\documentclass[%
 aip,
 amsmath,amssymb,
 reprint,
 floatfix,%
]{revtex4-1}

\usepackage{graphicx}
\usepackage{dcolumn}
\usepackage{bm}
\usepackage{float}
\usepackage[utf8]{inputenc}
\usepackage[T1]{fontenc}
\usepackage{mathptmx}
\usepackage{etoolbox}
\usepackage{subcaption}
\usepackage{enumitem}
\usepackage{booktabs}
\usepackage[colorlinks=true,linkcolor=blue,citecolor=blue,urlcolor=blue]{hyperref}
\usepackage[percent]{overpic}
\usepackage{xcolor}
\usepackage{comment}
\usepackage{booktabs}
\usepackage{makecell}
\usepackage{siunitx}
\usepackage{multirow}

\makeatletter
\def\@email#1#2{%
 \endgroup
 \patchcmd{\titleblock@produce}
 {\frontmatter@RRAPformat}
 {\frontmatter@RRAPformat{\produce@RRAP{*#1\href{mailto:#2}{#2}}}\frontmatter@RRAPformat}
 {}{}
}%
\makeatother

\begin{document}


\title[SCoReT: Super-Resolution of Turbulent Flows]
{SCoReT: Super-Resolution Compression and Reconstruction of Turbulent Flows}

\author{Royyuru Sai Prasanna Gangadhar}
\email{rspgangadharroyyuru@gmail.com}
\affiliation{
Indian Institute of Technology, Madras
}

\author{Shishir Srivastava}
\email{ic44283@imail.iitm.ac.in}
\affiliation{
Indian Institute of Technology, Madras
}

\author{Nagabhushana Rao Vadlamani}
\email{Corresponding author: nrv@smail.iitm.ac.in}
\affiliation{
Indian Institute of Technology, Madras
}

\author{Arvind Easwaran}
\email{arvinde@ntu.edu.sg}
\affiliation{
Nanyang Technological University, Singapore
}

\date{\today}

\begin{abstract}

High-fidelity simulations of the Navier--Stokes equations (NSE) generate massive amounts of data, motivating the need for efficient compression and reconstruction strategies for turbulent flows. At the same time, reconstructing flow fields from sparse measurements while retaining spectral content, turbulence statistics, and coherent structures remains a major challenge. This work investigates two complementary paradigms for turbulent flow reconstruction: supervised reconstruction and physics-informed reconstruction, in the context of transition to turbulence induced by three-dimensional distributed roughness elements. We introduce a vorticity-augmented supervised approach and a physics-informed approach, implemented through a partially assisted compressible PINN formulation based on the three-dimensional unsteady compressible Navier--Stokes equations. Reconstruction performance is evaluated at different sparsity levels using instantaneous velocity fields, mean-squared error, energy spectra, Reynolds stresses, turbulent kinetic energy, and Q-criterion isosurfaces. Rather than establishing a universal winner, the present study characterises the respective strengths, limitations, and operating regimes of these two approaches. The results indicate that at lower sparsity levels, the vorticity-augmented supervised model yields the lowest reconstruction error, recovers key statistical and spectral features, and enables substantial data compression. The PINN shows potential to reconstruct turbulent flows from sparse measurements without high-resolution labels and exhibits comparatively stable held-out extrapolation behaviour at higher sparsity. These results suggest the potential of combining data-driven and physics-informed learning for flow data compression and physics-informed reconstruction of turbulent flows from sparse data.

\end{abstract}

\maketitle


\section{\label{sec:Introduction}Introduction}
Turbulent flows exhibit a wide range of spatio-temporal scales, ranging from large coherent structures to small dissipative scales. Computational Fluid Dynamics (CFD) simulations solve the highly nonlinear Navier--Stokes equations (NSE) to resolve these turbulent structures. Such simulations are often computationally expensive and generate massive amounts of data, creating a need for efficient strategies for data storage, compression, and reconstruction. Early works have adopted modal decomposition techniques~\cite{taira2017modal,schmid2010dynamic} to identify dominant flow features and reduce the dimensionality of turbulent datasets. More recently, deep learning approaches have gained popularity by learning nonlinear mappings between input and output fields.

\begin{figure}[htbp]
  \centering
  \includegraphics[width = \linewidth]{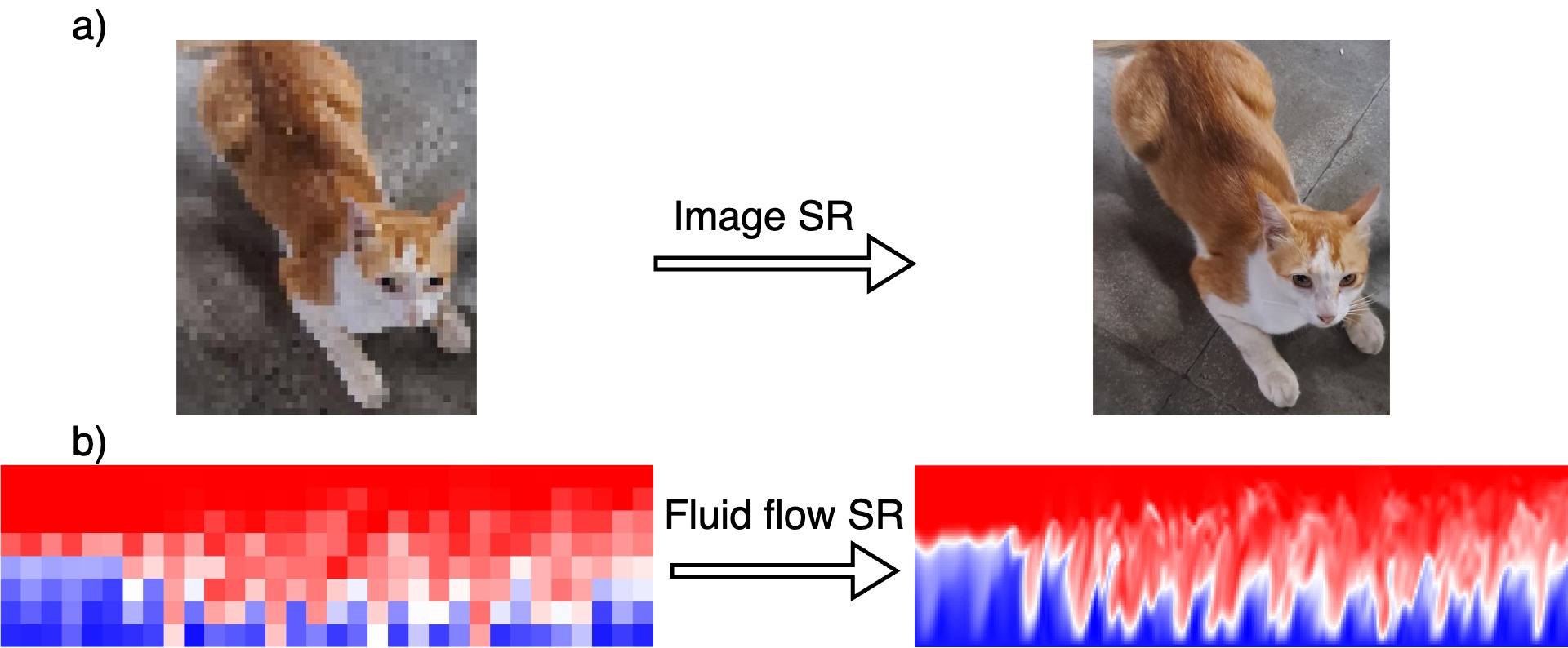}
  \caption{Figure illustrating the similarity between super-resolution of (a) general images and (b) fluid flows.}
  \label{fig:super_resol}
\end{figure}

Super-resolution~\cite{wang2020deep} (SR) is one such method that reconstructs a high-resolution field from a low-resolution input, recovering fine-scale information lost during downsampling. As illustrated in Figure~\ref{fig:super_resol}(a), a low-resolution image is provided as input and the model produces a super-resolved output with enhanced structural details. This concept can be extended to fluid flows as shown in Figure~\ref{fig:super_resol}(b), where a coarse flow field is provided as input to a model that reconstructs the corresponding high-resolution field. The reconstructed field aims to recover finer turbulent structures and near-wall gradients that are not present in the low-resolution input.

In the context of turbulent flows, SR-based reconstruction can be pursued through either supervised or physics-informed strategies. In the supervised case, SR enables flow compression and reconstruction, where only sparse or down-sampled fields are stored and the high-resolution flow is reconstructed during post-processing. In contrast, physics-informed reconstruction addresses the inverse problem of inferring the complete flow field from sparse measurements by embedding governing equations (e.g., NSE) into the learning process. 

Although both approaches operate on low-resolution or sparse flow information, they correspond to different practical objectives. In the supervised case, the low-resolution field is interpreted as compressed data from which the original high-resolution field is reconstructed, consistent with the classical super-resolution paradigm. In contrast, the physics-informed approach treats the retained low-resolution points as sparse measurements and seeks to infer the complete flow field without requiring paired high-resolution labels. In the present study, the same down-sampled datasets are used to investigate both perspectives, enabling a direct comparison between compression-oriented reconstruction and sparse-measurement-based flow reconstruction. A brief review of the relevant literature on supervised flow compression and physics-informed flow reconstruction is provided below.

\subsection{Flow compression strategies}
\label{subsection:flow_comp_lit_review}

Using supervised learning approaches, super-resolution methods have been widely applied to reconstruct 2D and 3D flow fields~\cite{fukami2019super,fukami2022machine,yousif2023deep,erichson2020shallow,liu2020deep,guemes2022super,wang2021novel,fukami2021global}. Fukami et al.~\cite{fukami2019super} introduced a hybrid Downsampled Skip-Connection Multi-Scale (DSC/MS) CNN architecture that improved robustness to rotation and translation (via DSC) while capturing turbulent structures across multiple scales (via MS). Wang et al.~\cite{wang2021novel} extended this idea with the Enhanced DSC/MS (E-DSC/MS) network, which uses a deeper architecture with smaller filters, merging multi-scale features and skip connections more effectively, and improved boundary continuity via reflection padding. Fukami et al.~\cite{fukami2021global} proposed a Voronoi-CNN for global flow reconstruction by projecting irregularly spaced sensor measurements onto a structured grid through Voronoi tessellation. Liu et al.~\cite{liu2020deep} introduced a Multi-temporal Path Convolutional Neural Network (MTPC) that takes a sequence of velocity snapshots as input and reconstructs the flow at the current timestep using multiple temporal paths, demonstrating improved recovery of spectral content compared to static single-frame CNNs. Few works have focused on full 3D reconstruction from reduced-dimensional inputs. Yousif et al.~\cite{yousif2023deep} proposed 2D-3DGAN, a generative adversarial network (GAN)-based framework combining 2D and 3D CNNs to reconstruct 3D turbulent velocity fields from intersecting planar inputs. Matsuo et al.~\cite{matsuo2024reconstructing} combined 2D and 3D CNN-based reconstruction with adaptive sampling for flow data compression, converting cross-sectional inputs into volumetric reconstructions. Erichson et al.~\cite{erichson2020shallow} used a shallow decoder that outperformed POD-based interpolation for flows past a cylinder and turbulent isotropic flow. Fukami et al.~\cite{fukami2022machine} applied 3D CNNs to reconstruct turbulent vortical structures in a pump sump from sparse surface pressure measurements.


\vspace{1em}

\subsection{Physics-informed flow reconstruction}
\label{subsection:pinn_lit_review}

Physics-informed neural networks (PINNs)~\cite{raissi2019physics} have emerged as a promising alternative to purely data-driven models for solving or reconstructing solutions to partial differential equations (PDEs), including steady and unsteady variants of the Navier--Stokes equations. While supervised approaches rely on high-resolution labels, PINNs incorporate physics-based loss functions to reconstruct the flow. Jin et al.~\cite{jin2021nsfnets} introduced NSFnets for incompressible NSE systems using velocity--pressure and velocity--vorticity formulations. Clark et al.~\cite{clark2018inferring} employed physics-informed spectral nudging for turbulence reconstruction and parameter inference under sparse measurements, with demonstrations on isotropic and rotating 3D turbulent flows. Sun et al.~\cite{sun2020physics} proposed a Bayesian physics-informed model (PC-BNN) for vascular flow reconstruction under sparse and noisy measurements, enabling uncertainty quantification. Several related frameworks have also been proposed for physics-informed super-resolution , including PhySR~\cite{ren2023physr}, MeshfreeFlowNet~\cite{esmaeilzadeh2020meshfreeflownet}, SURFNet~\cite{obiols2021surfnet}, and TransFlowNet~\cite{wang2022transflownet}. For irregular geometries and parametric PDEs, Gao et al.~\cite{gao2021phygeonet} proposed PhyGeoNet, while TPINN~\cite{manikkan2023transfer} introduced a domain-decomposition and parameter-sharing strategy for efficient learning. Additionally, Dwivedi et al.~\cite{dwivedi2020physics,dwivedi2022normal} proposed physics-informed extreme learning machines for fast solutions to linear PDEs, and FLRONet~\cite{dang2024flronet}, building on DeepONets~\cite{lu2021learning}, focused on flow reconstruction from sparse measurements with improved generalisation across CFD benchmark problems~\cite{luo2023cfdbench}.

Most supervised reconstruction studies have focused on canonical flows such as cylinder wakes, isotropic turbulence, and channel flows, whereas reconstruction performance in complex flow regimes that exhibit richer dynamics, three-dimensional flow structures, and temporally evolving compressible transitional turbulence phenomena remains underexplored. Likewise, many physics-informed approaches have largely been demonstrated for incompressible laminar or idealised turbulent flows with simplified boundary conditions, while compressible PINN formulations based on the three-dimensional, unsteady, and compressible Navier--Stokes equations remain under-explored for flow reconstruction. Consequently, the respective strengths and limitations of supervised reconstruction models and PINNs for such unsteady, three-dimensional, and compressible flow regimes remain unclear. Building on this motivation, the present work investigates the reconstruction of roughness-induced transitional turbulence using both supervised and physics-informed approaches. The objectives of this study are:

\begin{enumerate}
\item To evaluate supervised data-driven models and physics-informed neural networks (PINNs) as two complementary paradigms for the reconstruction of complex turbulent flows.
\item To validate the physical fidelity of the reconstructed flow fields using multi-scale statistical and spectral diagnostics, specifically focusing on Reynolds stress tensors, turbulent kinetic energy profiles, and spanwise energy spectra.
\item To evaluate how different levels of data compression affect reconstruction quality, and to rigorously test the ability of both AI paradigms to generalize across unseen spatial domains and future time steps.
\end{enumerate}

The objectives outlined above are pursued through several key contributions. The reconstruction problem is examined on a complex roughness-induced, three-dimensional transitional turbulent flow, extending beyond simplified canonical configurations commonly considered in prior studies. A vorticity augmented supervised objective and partially assisted compressible PINN formulation based on the three-dimensional, unsteady, and compressible Navier--Stokes equations are implemented for flow reconstruction. In addition, the respective behaviour of supervised and physics-informed models is characterized through spatial and spectral diagnostics, including Reynolds stresses and energy spectra. The goal is not to rank the two paradigms universally, but rather, to enable physically consistent reconstruction when high-resolution velocity labels are unavailable and to clarify the operating regimes in which each approach is most effective.

The remainder of the paper is organised as follows. Section~\ref{subsec:dataset} describes the dataset used in this study. Section~\ref{sec:flow_reconstruction} outlines the methodology, detailing the model architecture used. This section introduces the training objectives for supervised flow reconstruction, the physics-based loss functions for PINN, and the performance metrics used for evaluation. The key results and discussion are presented in Section~\ref{sec:results_discussion}. Finally, Section~\ref{sec:conclusion} summarises the conclusions and limitations of the present work and discusses potential future directions.

\section{Dataset and problem setup\label{subsec:dataset}}

\begin{figure}[t]
    \centering
    \includegraphics[width=\linewidth]{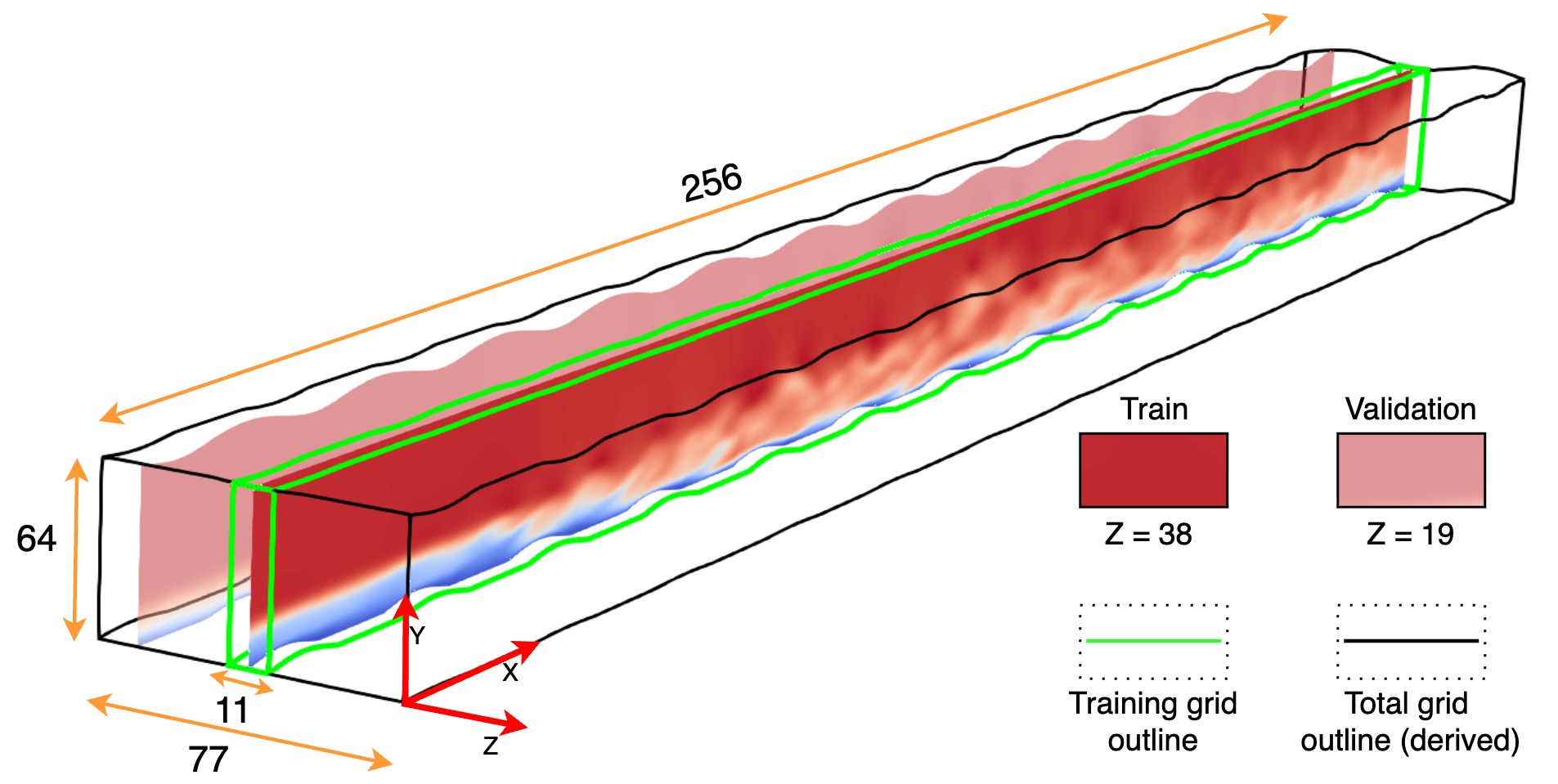}
    \caption{Computational grid extracted from the original dataset with streamwise velocity $u/U$ visualised to indicate transition to turbulence. The training grid, shown in green, spans $256 \times 64 \times 11$, while inference is performed over the full extracted domain of size $256 \times 64 \times 77$.}
    \label{fig:non_uniform_grid}
\end{figure}

\begin{figure}[t]
    \centering
    \includegraphics[width=\linewidth]{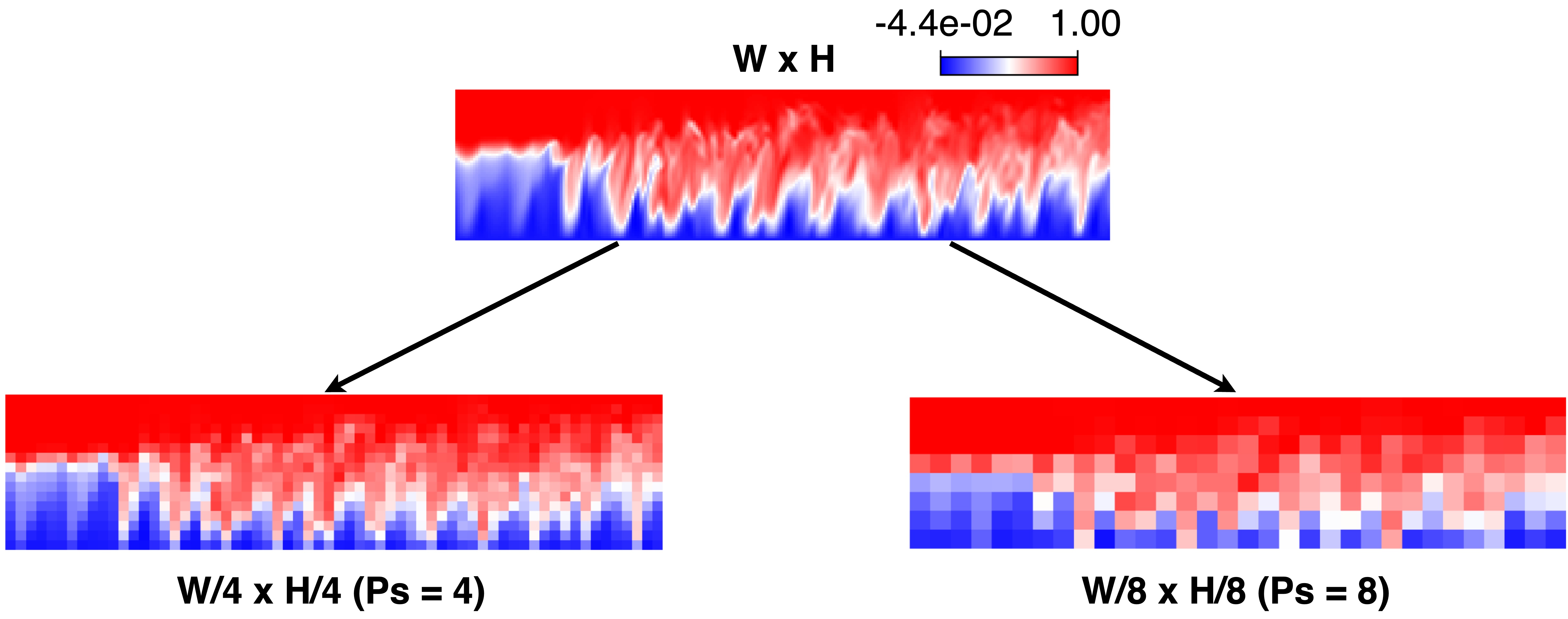}
    \caption{Illustration of LR data generated by down-sampling. The LR fields are generated by storing every 4th and 8th point along the $x$ and $y$ directions. These stored points may be interpreted either as compressed data or as sparse measurements.}
    \label{fig:super_resol_fluid_multi_ps}
\end{figure}


The dataset used in this study is adopted from Vadlamani et al.~\cite{vadlamani2018distributed}, which investigates the transition of a subsonic boundary layer over a flat plate with sinusoidal surface roughness distributed throughout the surface. The underlying scale-resolving simulation was performed on a non-uniform grid of size $516 \times 96 \times 96$ in the $(x,y,z)$ directions. 531 three-dimensional snapshots were stored at uniform non-dimensional time intervals of $\Delta t\,U/L = 0.0035$. Each snapshot contains the six primitive variables: density ($\rho$), pressure ($P$), temperature ($T$), and the three velocity components $(u,v,w)$ in the $(x,y,z)$ directions, respectively. The key flow parameters and dataset characteristics associated with the simulation are summarised in Table~\ref{tab:datasetdetails}.

To focus on the region where the flow transitions to turbulence and to reduce the computational cost associated with training on the entire dataset, a subzone of size $256 \times 64 \times 77$ is extracted from each snapshot. Figure~\ref{fig:non_uniform_grid} visualises this extracted computational subset (outlined in black) and the instantaneous contours of streamwise velocity $u/U$ to highlight the transition to turbulence. Thus, the final dataset consists of $531$ snapshots of the extracted subdomain, i.e.,
$531 \times 256 \times 64 \times 77$ values for each of the six primitive variables $(\rho,u,v,w,P,T)$. The original dataset is periodic in the $z$ direction. Since the dataset is sliced to a sub-region of $256 \times 64 \times 77$, we do not impose periodic boundary conditions along $z$ during any gradient calculations.

\begin{table}[h!]
\centering
\begin{tabular}{ll}
\toprule
\textbf{Parameter} & \textbf{Value} \\
\midrule
Reynolds number ($Re_L$) 
& 100{,}000 \\

Mach number ($M$) 
& 0.5 \\

Snapshot saving frequency 
& 50 time steps \\

Time interval between snapshots ($\Delta t\,U/L$) 
& 0.0035 \\

Total number of snapshots 
& 531 \\
\bottomrule
\end{tabular}
\caption{\label{tab:datasetdetails}Key flow parameters and dataset details.}
\end{table}

The training dataset consists of the first 425 snapshots (0 - 424) and a restricted spanwise region containing 11 planes in the $z$-direction (highlighted in green in Figure~\ref{fig:non_uniform_grid}). This subdomain, referred to as the \emph{training grid}, has a size of $256 \times 64 \times 11$. Thus, all models are trained using a total of
$425 \times 256 \times 64 \times 11$ spatio-temporal samples. The remaining snapshots and spanwise planes form the validation dataset, which is used to evaluate the held-out extrapolation behaviour of the reconstruction models over both unseen snapshots $t \in  \{425,\dots, 530\}$ and unseen spanwise planes $z \in \{0, \dots, 32\} \cup \{44, \dots, 76\}$. The low-resolution (LR) inputs are generated by spatial down-sampling of the high-resolution (HR) data. Specifically, every $P_s$-th point is stored along the streamwise and wall-normal directions $(x,y)$, while retaining the full spanwise resolution in $z$. Starting from the extracted HR subdomain of size $(256 \times 64 \times 77)$, the corresponding LR field has the size $(256/P_s) \times (64/P_s) \times 77$.

In this study, we consider $P_s = 4$ and $8$. The down-sampled LR data can be interpreted in two equivalent ways depending on the application: (i) as \emph{compressed data} stored to disk for later reconstruction, or (ii) as \emph{sparse measurements} available for inverse complete flow reconstruction. Figure~\ref{fig:super_resol_fluid_multi_ps} illustrates representative HR--LR pairs for the considered values of $P_s$. As the compression level increases from $P_s=4$ to $P_s=8$, progressively larger fractions of the original spatial information are removed, leading to a coarser representation of the underlying flow structures in the LR input. This increasing loss of resolved spatial content makes the reconstruction problem more challenging and provides a controlled basis for assessing model robustness under different compression levels.


\section{Methodology\label{sec:flow_reconstruction}}
Building on the dataset described in Section~\ref{subsec:dataset}, this section outlines two training strategies for reconstructing flow: i) A vorticity-augmented supervised loss, which penalizes the L2 norm of both velocity and vorticity (LVV), and  ii) A physics-informed neural network (PINN) formulation that reconstructs the flow by enforcing the three-dimensional compressible and unsteady Navier-Stokes equations (NSE).  The model architecture used for both approaches is discussed in Section \ref{subsec:model_description}, while a detailed discussion of LVV and PINN can be found in Sections \ref{subsec:LVV} and \ref{subsec:pinn_reconstruction}, respectively.

\subsection{Model architecture \label{subsec:model_description}}
\begin{figure*}
    \raggedright
    \includegraphics[width=0.9\textwidth]{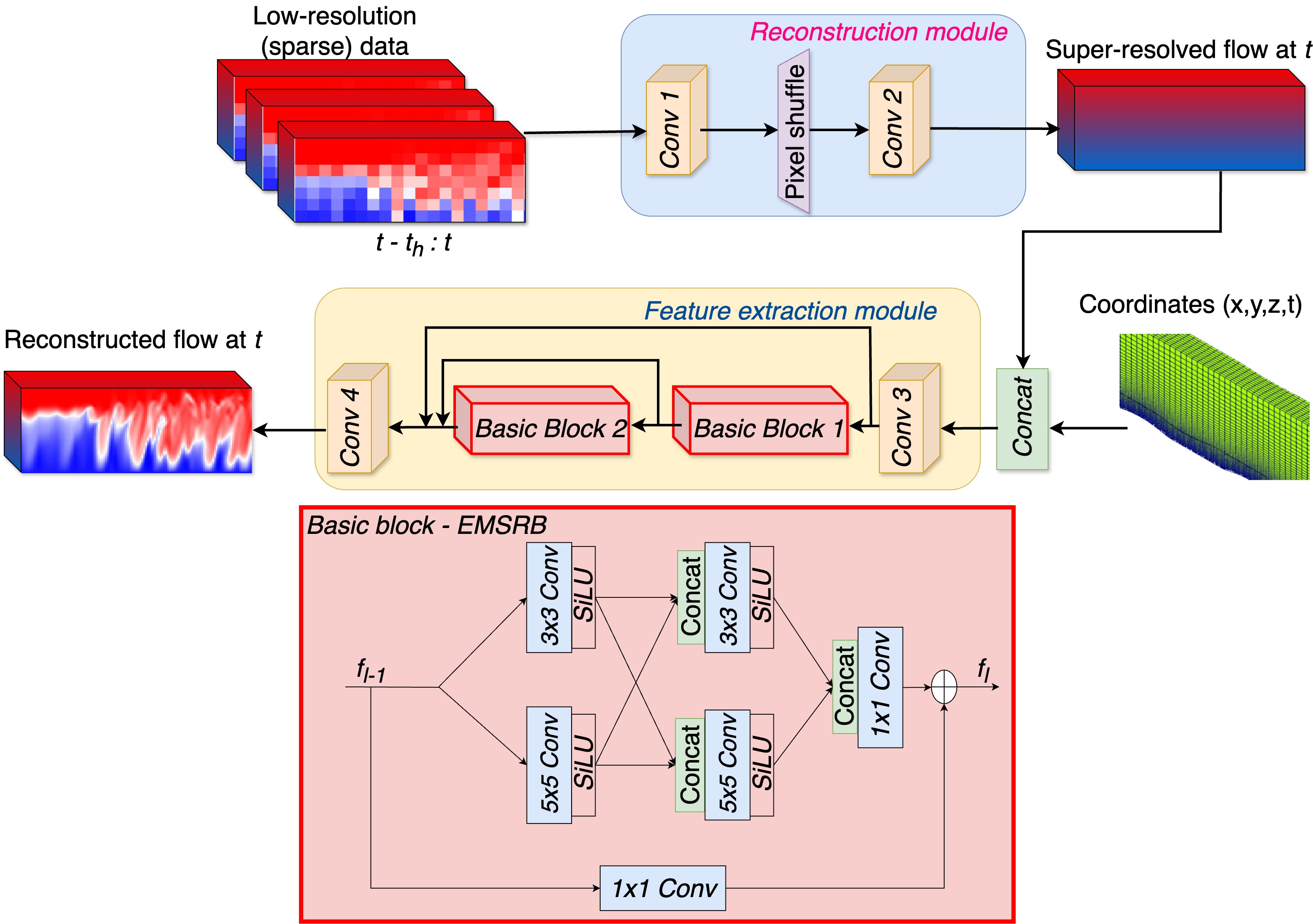}
    \caption{Model architecture used for supervised and physics-informed reconstruction. The reconstruction module (blue) up-samples the sparse input to the training-grid resolution, while the feature extraction module (yellow) learns flow features to reconstruct fluid flow. Figure adapted from Liu et al. \cite{liu2020deep}.}
    \label{fig:btpmodel}
\end{figure*}
The model employed in this study is based on the Multi-Temporal Path CNN (MTPC) architecture introduced by Liu et al.\cite{liu2020deep}. MTPC was selected for its incorporation of the Enhanced Multi-Scale Residual Block (EMSRB), which has demonstrated strong capability in extracting features across multiple scales. This makes the architecture particularly well-suited for modelling turbulent flows, which are characterised by several spatiotemporal scales. The original MTPC utilises previous and future snapshots through a combination of Back Temporal Path (BTP), Centre Temporal Path (CTP), and Forward Temporal Path (FTP), equivalent to backward, central and forward differential schemes, respectively, in CFD. Nevertheless, time-marching in most CFD solvers typically uses a backwards differencing scheme to predict the flow at the next time step. Hence, the current approach utilises only the BTP. Another important modification is the extension of the BTP model to reconstruct three-dimensional turbulent flow fields rather than the two-dimensional unsteady flow reconstructed by Liu et al. \cite{liu2020deep}. Figure~\ref{fig:btpmodel} provides a schematic of the model, which consists of two sequential submodules: 1) Reconstruction module (RM) and 2) Feature extraction module (FEM), which together predict the super-resolved flow. For the remainder of this section, $b_s$ is the batch size, $t_h$ is the number of previous downsampled snapshots provided to the model as input, $c$ is the number of variables reconstructed, $P_s$ is the upsampling factor, and $n_x,n_y,n_z$ represent the total number of grid points present along $x,y,z$ directions, respectively. A brief description of these modules is given below:

\begin{enumerate}
    \item \textbf{Reconstruction module (RM) \text:} 
    The low-resolution input of shape $(b_s, t_h, c, n_z, n_y/P_s, n_x/P_s)$ is reshaped to $(b_s, t_h \times c \times n_z, n_y/P_s, n_x/P_s)$ and passed through the reconstruction module to generate an upsampled output of shape $(b_s, c, n_z, n_y, n_x)$. 
    The initial convolution \texttt{Conv1} produces a feature tensor of shape $(b_s, c \times n_z \times P_s^2, n_y/P_s, n_x/P_s)$, which is then rearranged by an upsampling layer (Pixel shuffle) to $(b_s, c \times n_z, n_y, n_x)$. Pixel shuffle, introduced by Shi et al.~\cite{shi2016real}, is an efficient sub-pixel upsampling operation that rearranges the tensor from shape 
    \((b_s, c \times n_z \times P_s^2, {n_y}/{P_s}, {n_x}/{P_s})\) 
    to  \((b_s, c \times n_z, n_y.P_s/P_s, n_x.P_s/P_s)\), 
    effectively increasing the spatial resolution by a factor of \(P_s\). A second convolution \texttt{Conv2} retains this upsampled resolution, and the tensor is reshaped to $(b_s, c, n_z, n_y, n_x)$ for downstream processing. 
    This module is required to align the dimensions of the predicted velocity field with the training grid.

    \item \textbf{Feature extraction module (FEM):} 
    The upsampled tensor is concatenated along the channel dimension with spatio-temporal coordinates to form an input of shape $(b_s, c + coords_{(x,y,z,t)}, n_z, n_y, n_x)$. The concatenated tensor is reshaped to $(b_s, (c + coords_{(x,y,z,t)}) \times n_z, n_y, n_x)$ and passed through \texttt{Conv3} to extract feature maps of shape $(b_s, f, n_y, n_x)$. These are processed through two sequential basic blocks (EMSRB) to capture multi-scale flow features, and skip connections are employed to enhance feature representation. An inset plot in Figure \ref{fig:btpmodel} illustrates the enhanced multi-scale residual block (EMSRB) block. The EMSRB, proposed by Liu et al. \cite{liu2020deep}, serves as the basic block in FEM and extends the Multi-scale Residual Block (MSRB \cite{li2018multi}). As shown in Figure \ref{fig:btpmodel}, two sequential EMSRB blocks are used in the FEM to extract flow features. Convolutional kernels of different sizes are used to extract flow features at various scales. For example, the $3 \times 3$ kernel extracts smaller scales while the $5 \times 5$ kernel extracts relatively larger scales. These features are further cross-concatenated and convoluted to facilitate effective feature interactions. The resulting tensor of shape $(b_s, 3 \times f, n_y, n_x)$ is mapped through \texttt{Conv4} to an output tensor of shape $(b_s, c \times n_z, n_y, n_x)$ and reshaped to $(b_s, c, n_z, n_y, n_x)$ for downstream processing.
\end{enumerate}
The activation function used in the model is the Sigmoid Linear Unit (SiLU) in FEM. The mathematical form of the SiLU activation is defined as SiLU$(x)=x/(1+e^{-x})$. SiLU is used in FEM because its smooth, differentiable form is advantageous when differentiating the network during the backward pass for physics-informed training. We utilise five previous snapshots ($t_h$) of low-resolution (LR) data as input to RM. The batch size is set at 4, and the learning rate is fixed at \( 10^{-4} \). The principal architectural and training hyperparameters, including activation functions, are summarised in Table~\ref{tab:btpmodel}.

\begin{table}[h!]
\centering
\begin{tabular}{ll}
\toprule
\textbf{Parameter} & \textbf{Value} \\
\midrule

\multicolumn{2}{c}{\textit{Input configuration}} \\
\midrule
$n_x,n_y,n_z$ & $256,64,11$ \\
Number of previous snapshots ($t_h$) & 5 \\
Batch size ($b_s$) & 4 \\
Channels ($c$) & $3$ ($u,v,w$) \\
Upsampling factor ($P_s$) & $4, 8$ \\
Input shape & $(b_s, t_h, c, n_z, n_y/P_s, n_x/P_s)$ \\

\midrule
\multicolumn{2}{c}{\textit{Reconstruction module (RM)}} \\
\midrule
Reshaped input & $(b_s, t_h \cdot c \cdot n_z, n_y/P_s, n_x/P_s)$ \\
Conv1 channels & $c \cdot n_z \cdot P_s^2$ \\
Upsampling & Pixel shuffle ($\times P_s$) \\
Conv2 channels & $c \cdot n_z$ \\
RM output shape & $(b_s, c, n_z, n_y, n_x)$ \\

\midrule
\multicolumn{2}{c}{\textit{Feature Extraction Module (FEM)}} \\
\midrule
Input with coordinates & $(b_s, c + 4, n_z, n_y, n_x)$ \\
Reshaped input & $(b_s, (c+4)\cdot n_z, n_y, n_x)$ \\
Conv3 channels & $f = 64$ \\
Basic block & EMSRB  \\
Kernel sizes in EMSRB & $3\times3$, $5\times5$ \\
Number of EMSRB blocks & 2 (sequential) \\
Conv4 channels & $c \cdot n_z$ \\
FEM output shape & $(b_s, c, n_z, n_y, n_x)$ \\
Activation  & SiLU \\

\midrule
\multicolumn{2}{c}{\textit{Training parameters}} \\
\midrule
Learning rate & $10^{-4}$ \\
\bottomrule
\end{tabular}
\caption{Architectural and training hyperparameters of the BTP model used for 3D flow reconstruction.}
\label{tab:btpmodel}
\end{table}

With the model architecture established, the training strategies based on vorticity-augmented supervision and physics-informed learning are described next.

\subsection{LVV: Vorticity-augmented supervised compression}
\label{subsec:LVV}

The standard velocity-only MSE loss does not explicitly emphasize regions of strong shear and rotational motion, which are critical in turbulent flows. To improve reconstruction fidelity in such regions, we augment the supervised objective with a vorticity-based penalty. This approach, termed LVV, balances the reconstruction of both the velocity field $\mathbf{f} = [u, v, w]^T$ and its corresponding curl, the vorticity field $\boldsymbol{\omega} = \nabla \times \mathbf{f} = [\omega_x, \omega_y, \omega_z]^T$. True and predicted vorticity fields are computed via an explicit fourth-order finite-difference (E4) scheme. Thus, the total training loss for LVV (as described in equation~\ref{eq:loss_LVV}) is the sum of the mean squared error (MSE) of velocity ($J_{\text{vel}}$) and the normalized mean squared error of vorticity ($J_{\text{vort}}$) as :

\begin{equation}
J_{\text{LVV}} = J_{\text{vel}} + J_{\text{vort}}
\label{eq:loss_LVV}
\end{equation}

where both terms compute an element-wise average over the batch size $b_s$, the total spatial grid points per sample $N_s$ ($N_s = n_x \times n_y \times n_z$), and the $c=3$ vector components ($c \in \{x, y, z\}$):

\begin{equation}
J_{\text{vel}} = \frac{1}{ b_s N_sc} \sum_{b=1}^{b_s} \sum_{i=1}^{N_s} \sum_{c=1}^{3} \left( \hat{f}_{b,i,c} - f_{b,i,c} \right)^2
\end{equation}

\begin{equation}
J_{\text{vort}} = \frac{1}{b_s N_sc} \sum_{b=1}^{b_s} \sum_{i=1}^{N_s} \sum_{c=1}^{3} \left( \tilde{\omega}_{b,i,c} \right)^2
\end{equation}

Here, $f_{b,i,c}$ and $\hat{f}_{b,i,c}$ represent the component $c$ of the true and predicted velocity vectors at spatial location $i$ of the $b$-th batch element. Since vorticity can take very large values, to maintain numerical stability during training, the normalised vorticity error component $\tilde{\omega}_{b,i,c}$ is scaled using a spatial maximum absolute value calculated independently for each batch element and component as:

\begin{equation}
\tilde{\omega}_{b,i,c} = \frac{\omega_{b,i,c}(\hat{\mathbf{f}}) - \omega_{b,i,c}(\mathbf{f})}{\max_{j \in \{1, \dots, N_s\}} \left| \omega_{b,j,c}(\hat{\mathbf{f}}) - \omega_{b,j,c}(\mathbf{f}) \right| + \epsilon}
\end{equation}

where $\epsilon = 10^{-16}$ is introduced to prevent division by zero.
\subsection{Physics-informed flow reconstruction using PINNs\label{subsec:pinn_reconstruction}}
The supervised reconstruction strategy, LVV, introduced above, relies on paired low and high-resolution velocity fields during training. In many practical applications, however, high-resolution labels may not be available, and only sparse flow measurements may be accessible. To address this scenario, we also evaluate a physics-informed reconstruction method. PINNs embed the governing physical equations directly into the learning objective by minimising the NSE residual while simultaneously enforcing consistency with available sparse measurements, boundary and initial conditions. In the present study, we evaluate a partially assisted compressible PINN formulation in which the three-dimensional unsteady, compressible Navier--Stokes residual is enforced for reconstructing a transitional boundary-layer flow $(u,v,w)$, while the thermodynamic variables $(\rho,P,T)$ required in the residual are supplied from the reference dataset~\cite{vadlamani2018distributed}. This allows the present study to focus specifically on velocity-field reconstruction while retaining physical consistency through the compressible and unsteady NSE residuals. We highlight that, unlike LVV, PINN works on a more stringent constraint and does not use paired LR-HR data. This formulation, based on the compressible, unsteady three-dimensional NSE, remains relatively underexplored in the existing PINN-based turbulence reconstruction literature. Extension of the framework to reconstruct all six primitive variables is left for future work. 

The network parameters are optimised by minimising a composite loss that enforces (i) consistency with the governing NSE equations through a PDE-residual term, and (ii) agreement with the available information in the form of sparse measurements, boundary conditions, and the initial condition. These loss components are detailed below

\subsubsection{Navier--Stokes residual loss\label{subsec:NVloss}}
In the PINN formulation, physical consistency is enforced by augmenting the training objective with an NSE residual loss, constructed from the governing equations of three-dimensional compressible and unsteady flow. Specifically, the network prediction is constrained to satisfy the conservation laws of mass, momentum, and total energy via the residuals of the compressible, unsteady NSE equations. To evaluate these residuals during training, the coordinate tensor (passed as input to FEM, refer to figure~\ref{fig:btpmodel}) is marked with \texttt{requires\_grad=True}, allowing the derivatives required for the PDE residual to be computed through automatic differentiation.
The governing three-dimensional compressible and unsteady Navier--Stokes equations in Cartesian coordinates are written as
\begin{align}
\frac{\partial Q}{\partial t} + 
\frac{\partial E}{\partial x} + 
\frac{\partial F}{\partial y} + 
\frac{\partial G}{\partial z} =
\frac{\partial E_{v}}{\partial x} + 
\frac{\partial F_{v}}{\partial y} + 
\frac{\partial G_{v}}{\partial z},
\label{eq:compressible_NS}
\end{align}
where $Q$ is the conserved variable vector and $(E,F,G,E_v,F_v,G_v)$ are the flux vectors. Please refer to Appendix \ref{subsec:appendix_nvlos} for more details on NSE loss.

\subsubsection{Known-data, boundary-condition, and initial-condition losses\label{subsec:known_data_loss}}

In addition to the PDE residual loss (Section~\ref{subsec:NVloss}), the PINN training objective includes losses associated with known sparse measurements, boundary conditions, and the initial condition. The known-data loss penalizes mismatch between reconstructed and reference velocities at the retained LR locations (i.e., every $P_s$-th point in the $(x,y)$ directions), the initial-condition loss enforces agreement at the first snapshot, and the boundary-condition loss imposes consistency on all six boundary planes of the training grid.

The total PINN loss for the true velocity field $\mathbf{f} = [u,v,w]^T$ and predicted velocity field $\mathbf{\hat{f}}$ is expressed as a weighted sum in equation~\ref{eq:pinnloss} as,
\begin{equation}
J_{\mathrm{PINN}} = \lambda_{pde} J_{pde} + \lambda_{bc} J_{bc} + \lambda_{ic} J_{ic} + \lambda_{known} J_{known},
\label{eq:pinnloss}
\end{equation}
where
\begin{equation}
\begin{split}
J_{pde} &= \frac{1}{b_s N_{pde}}
\sum_{b=1}^{b_s}\sum_{i=1}^{N_{pde}}
\left\|\mathbf{\hat{g}}(x_{pde}^{b,i},t_{pde}^{b,i})\right\|_2^2, \\
&\qquad \text{where } N_{pde}
= n_x \times n_y \times n_z
\end{split}
\label{eq:losspde}
\end{equation}

\begin{equation}
\begin{split}
J_{bc} &= \frac{1}{b_s N_{bc}c}
\sum_{b=1}^{b_s}\sum_{i=1}^{N_{bc}}
\left\|\hat{\mathbf{f}}(x_{bc}^{b,i},t_{bc}^{b,i})
-\mathbf{f}(x_{bc}^{b,i},t_{bc}^{b,i})\right\|_2^2, \\
&\qquad \text{where } N_{bc}
=2(n_x n_y+n_y n_z+n_x n_z)
\end{split}
\label{eq:loss_bc}
\end{equation}

\begin{equation}
\begin{split}
J_{ic} &= \frac{1}{N_{ic}c}
\sum_{i=1}^{N_{ic}}
\left\|\hat{\mathbf{f}}(x_{ic}^{i},0)
-\mathbf{f}(x_{ic}^{i},0)\right\|_2^2, \\
&\qquad \text{where } N_{ic}
= n_x \times n_y \times n_z
\end{split}
\label{eq:lossic}
\end{equation}

\begin{equation}
\begin{split}
J_{known} &= \frac{1}{b_s N_{known}c}
\sum_{b=1}^{b_s}\sum_{i=1}^{N_{known}}
\left\|\hat{\mathbf{f}}(x_{known}^{b,i},t_{known}^{b,i})
-\mathbf{f}(x_{known}^{b,i},t_{known}^{b,i})\right\|_2^2, \\
&\qquad \text{where } N_{known}
=\frac{n_x}{P_s}\times\frac{n_y}{P_s}\times n_z
\end{split}
\label{eq:knowndataloss}
\end{equation}
Here, $\mathbf{\hat{g}}$ denotes the vector of Navier--Stokes residuals evaluated from the network prediction and $c = 3$ denotes the three velocity components. The results presented in this work are obtained using $\lambda_{pde}=0.5$, and $\lambda_{bc}=\lambda_{ic}=\lambda_{known}= 1$, and the weight terms are normalised to a magnitude of 1.0. 


\subsection{Metrics used to evaluate models}
Model performance is evaluated using both error-based and physics-based diagnostics, including mean squared error, Reynolds stresses, turbulent kinetic energy, and energy spectra. While mean squared error provides a global measure of reconstruction accuracy, it may not fully capture local turbulent dynamics. For this reason, statistical diagnostics such as Reynolds stresses and turbulent kinetic energy are examined to assess the extent to which key turbulent statistics are recovered in the reconstructed flow. In addition, because turbulent flows contain structures spanning multiple scales, energy spectra are analysed to assess how well the reconstructed fields capture scale-dependent behaviour and to characterize model behaviour across the range of resolved turbulent scales.

\section{Results and discussion\label{sec:results_discussion}}
The reconstruction performance of the LVV model and the PINN is assessed using instantaneous flow fields, quantitative error measures, spectral diagnostics, and turbulent statistics. It is reiterated that models are trained on a subdomain of size \(425\times256\times64\times11\), while inference is performed on the full domain of size \(531\times256\times64\times77\), enabling evaluation both within and beyond the training region. Two planes, \(z=38\) and \(z=19\) (refer to Figure \ref{fig:non_uniform_grid}), are selected as representative planes within the training grid and outside the training grid, respectively. The performance at \(z=38\) during the first 425 snapshots is used to evaluate reconstruction accuracy within the training data. In contrast, the results at \(z=38\) from the last 106 snapshots are utilised to assess temporal generalisation. The performance on \(z=19\) with the training snapshots provides a measure of spatial generalisation, while the performance over \(z=19\) during the last 106 snapshots offers insight into spatial-temporal generalisation. LVV is trained for 1000 epochs, while PINN is trained for 100 epochs, after which no significant reduction in loss is observed for PINN. All assessments follow the same post-processing, which includes 3D Gaussian smoothing with a standard deviation of 1.0 along each axis, applied to both the DNS and reconstructed fields. We first examine instantaneous field reconstruction and quantitative error analysis, followed by spectral fidelity and turbulent statistics, through which the effects of compression and out-of-training-domain behaviour are characterized.

\subsection{Instantaneous velocity reconstruction\label{subsec:inst_vel}}
\begin{figure*}[!htbp]
\centering
    \begin{subfigure}{0.48\textwidth}
        \centering
        \includegraphics[width=\textwidth]{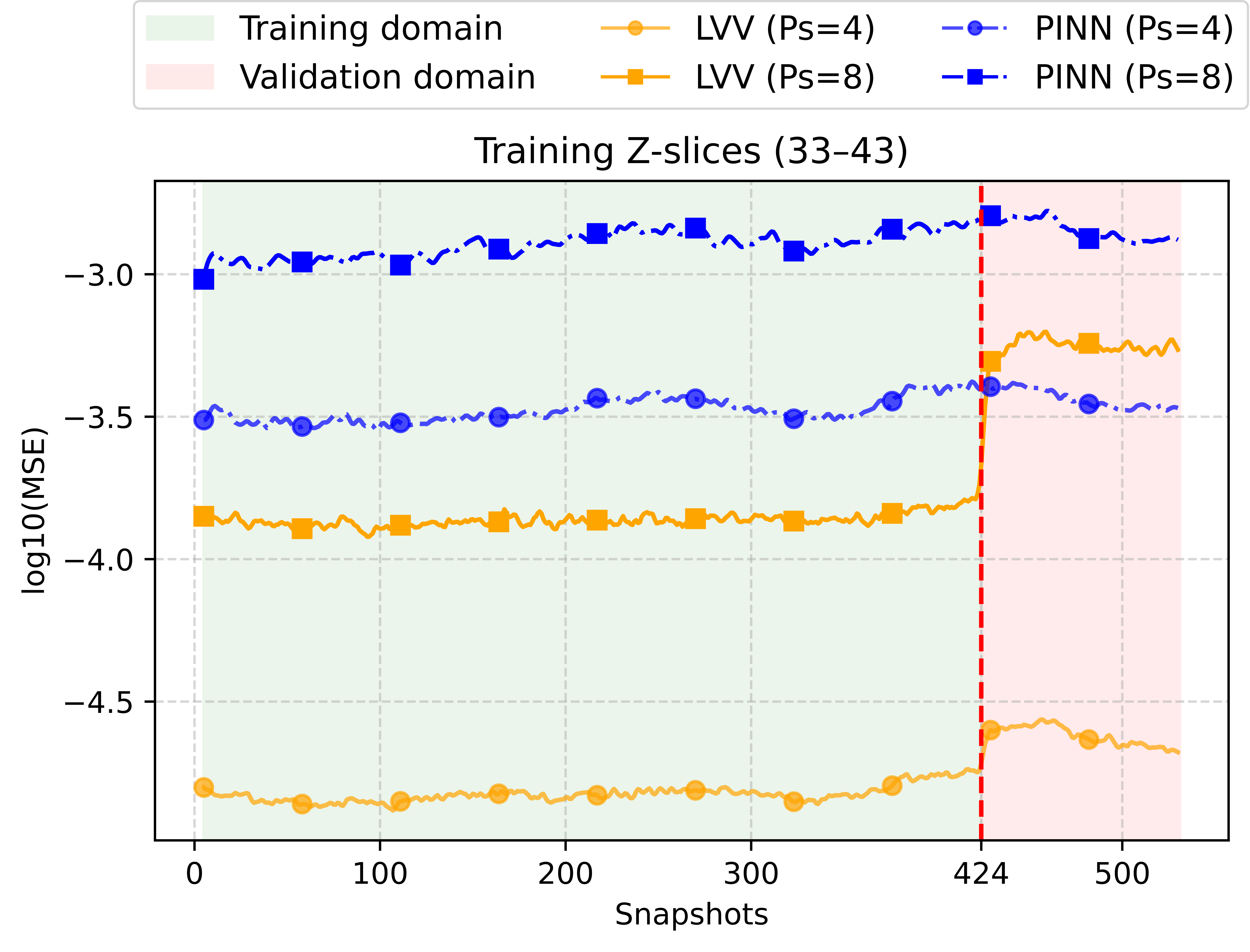}
        \caption{Training $z$-slices}
        \label{fig:train_z_slices}
    \end{subfigure}
    \hfill 
    \begin{subfigure}{0.48\textwidth}
        \centering
        \includegraphics[width=\textwidth]{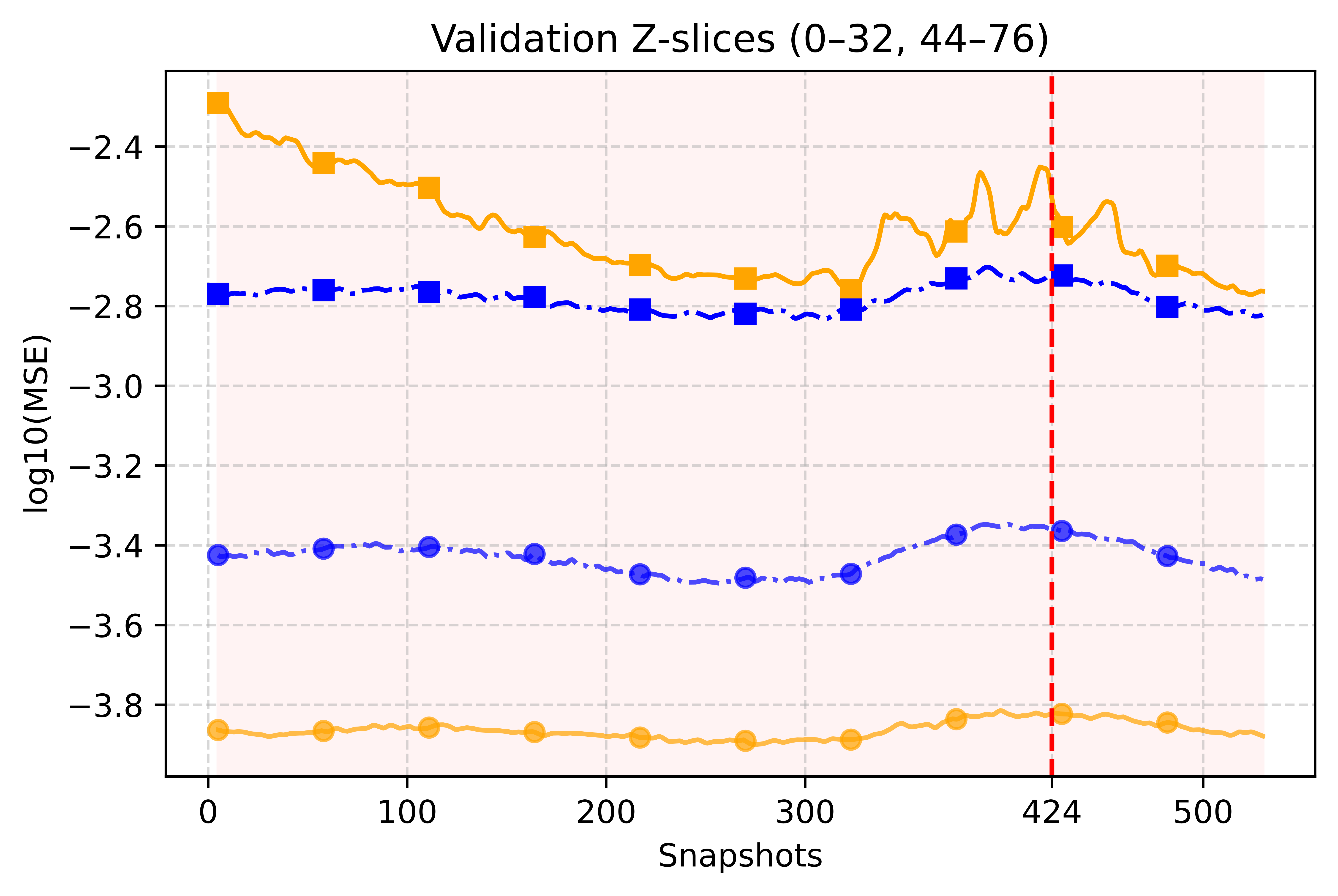}
        \caption{Validation $z$-slices}
        \label{fig:val_z_slices}
    \end{subfigure}

    \vspace{0.2cm} 

    \begin{subfigure}{0.48\textwidth}
        \centering
        \includegraphics[width=\textwidth]{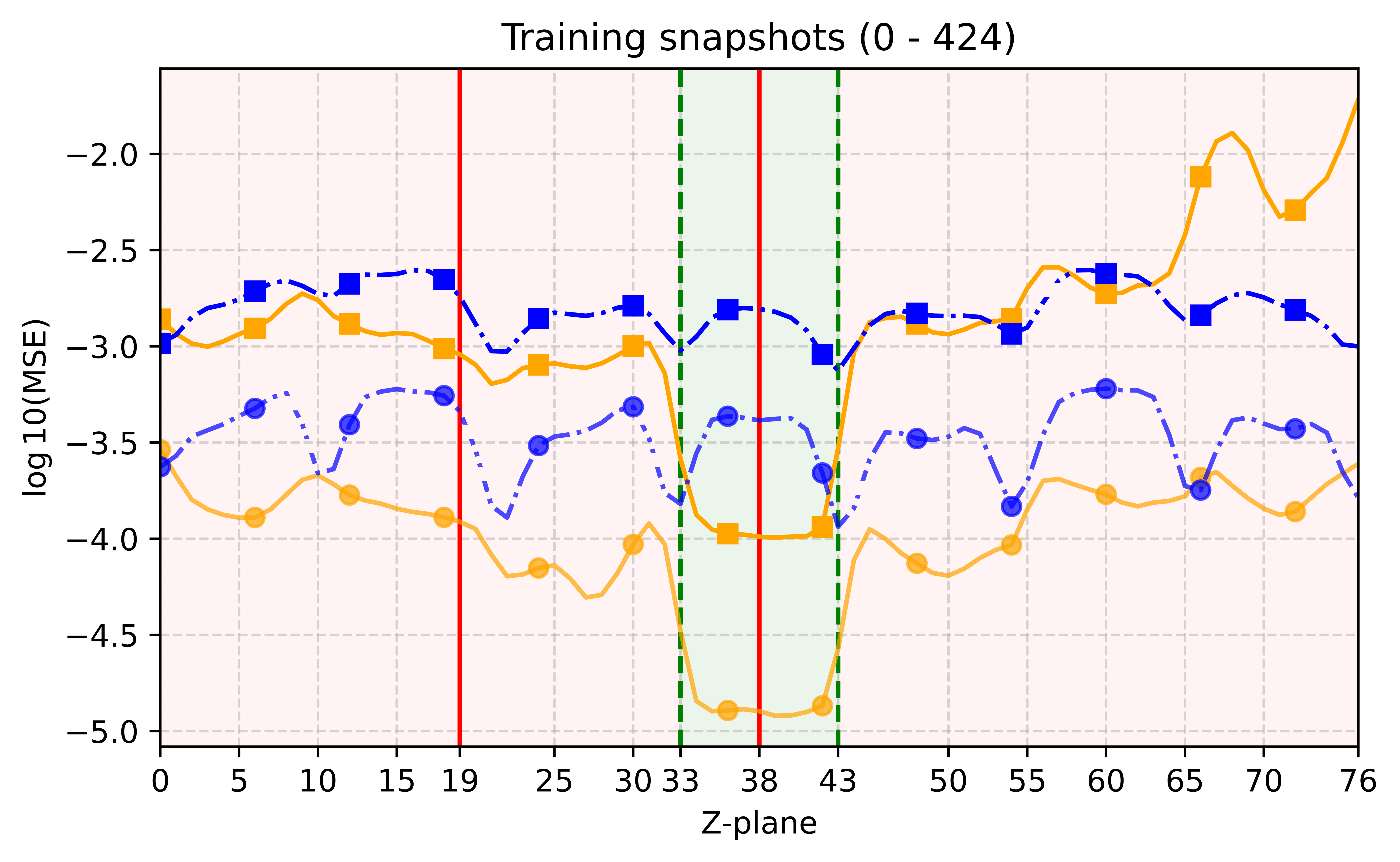}
        \caption{Training Snapshots}
        \label{fig:train_snapshots}
    \end{subfigure}
    \hfill
    \begin{subfigure}{0.48\textwidth}
        \centering
        \includegraphics[width=\textwidth]{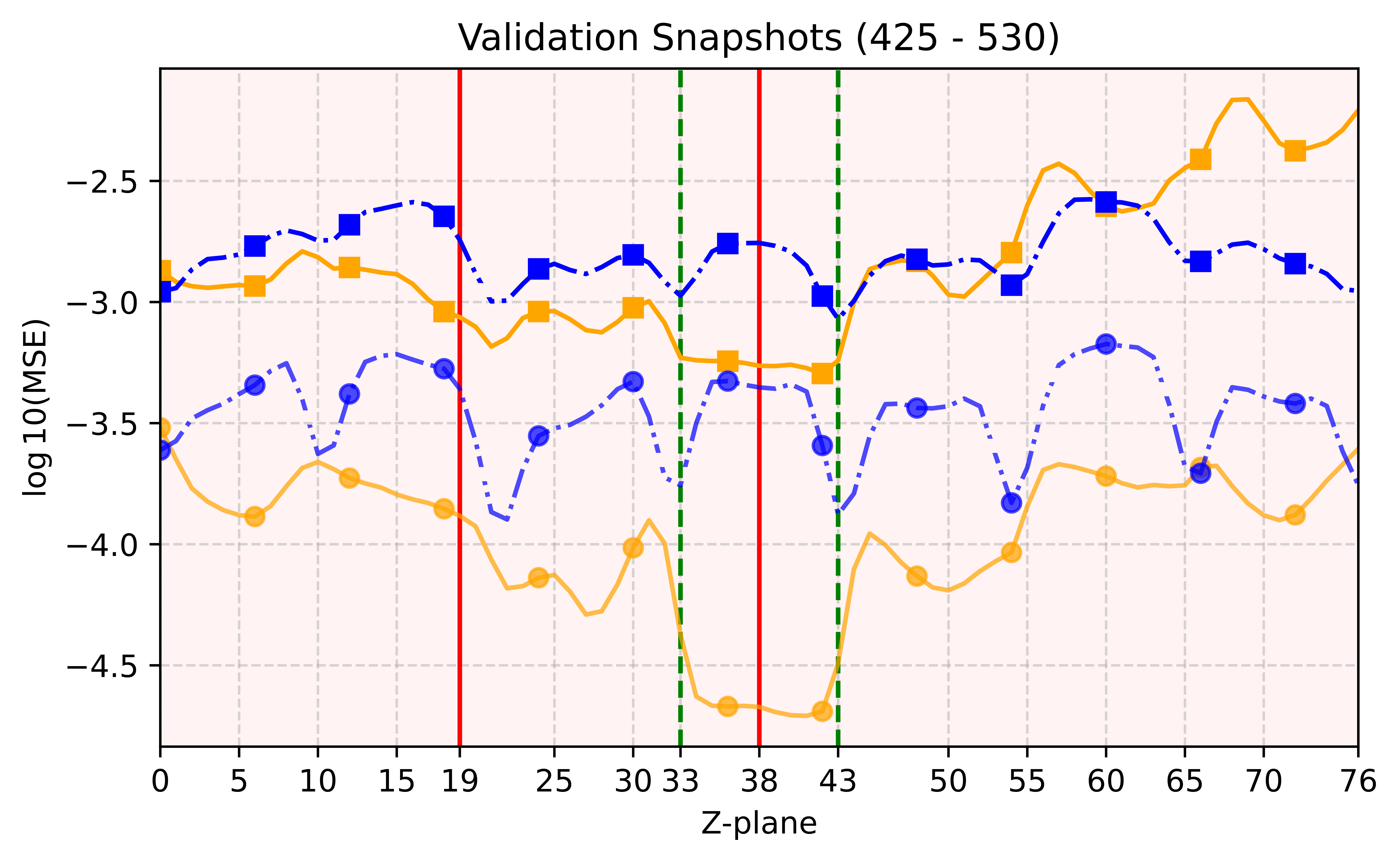}
        \caption{Validation Snapshots}
        \label{fig:val_snapshots}
    \end{subfigure}

\caption{Comprehensive Mean Squared Error (MSE) reconstruction analysis for LVV and PINN. Panels (a) and (b) show instantaneous MSE averaged across spanwise locations for training and validation $z$-slices. Panels (c) and (d) show MSE along spanwise $z$-planes averaged across training and validation snapshots. The shaded regions indicate the training (green) and validation (red) domains.}
\label{fig:mse_line_plot}
\end{figure*}

\sisetup{
  scientific-notation = true,
  round-mode = figures,
  round-precision = 3,
  detect-all,
  input-exponent-markers = {e,E}
}
\sisetup{
  scientific-notation = true,
  round-mode = figures,
  round-precision = 3,
  detect-all,
  input-exponent-markers = {e,E}
}

\begin{table*}[t]
\centering
\small
\begin{tabular}{@{}lllcSSSS@{}}
\toprule
\textbf{Domain} & \textbf{Channel} & \textbf{Snapshot} &
\multicolumn{2}{c}{$P_s=4$} &
\multicolumn{2}{c}{$P_s=8$} \\
\cmidrule(lr){4-5} \cmidrule(lr){6-7}

& & &
\multicolumn{1}{c}{PINN} &
\multicolumn{1}{c}{LVV} &
\multicolumn{1}{c}{PINN} &
\multicolumn{1}{c}{LVV} \\
\midrule

\multirow{6}{*}{Across 77 planes}
& \multirow{2}{*}{$u$} & 212 & $9.46\times10^{-4}$ & 4.30e-04 & 4.46e-03 & 5.71e-03 \\
& & 478 & $1.06\times10^{-3}$ & 4.42e-04 & 4.68e-03 & 5.94e-03 \\
\cmidrule(lr){2-7}

& \multirow{2}{*}{$v$} & 212 & $2.00\times10^{-1}$ & 2.98e-02 & 6.63e-01 & 5.45e-01 \\
& & 478 & $2.04\times10^{-1}$ & 3.22e-02 & 6.75e-01 & 4.52e-01 \\
\cmidrule(lr){2-7}

& \multirow{2}{*}{$w$} & 212 & $1.38\times10^{-1}$ & 3.18e-02 & 6.18e-01 & 6.93e-01 \\
& & 478 & $1.45\times10^{-1}$ & 3.60e-02 & 6.20e-01 & 5.82e-01 \\
\midrule

\multirow{6}{*}{Plane $z=19$}
& \multirow{2}{*}{$u$} & 212 & $1.41\times10^{-3}$ & 4.56e-04 & 5.18e-03 & 2.92e-03 \\
& & 478 & $1.39\times10^{-3}$ & 4.73e-04 & 6.05e-03 & 3.04e-03 \\
\cmidrule(lr){2-7}

& \multirow{2}{*}{$v$} & 212 & $2.27\times10^{-1}$ & 1.45e-02 & 6.75e-01 & 2.04e-01 \\
& & 478 & $1.80\times10^{-1}$ & 1.72e-02 & 6.58e-01 & 2.34e-01 \\
\cmidrule(lr){2-7}

& \multirow{2}{*}{$w$} & 212 & $1.35\times10^{-1}$ & 3.22e-02 & 6.23e-01 & 3.17e-01 \\
& & 478 & $1.57\times10^{-1}$ & 3.45e-02 & 6.50e-01 & 3.85e-01 \\
\midrule

\multirow{6}{*}{Plane $z=38$}
& \multirow{2}{*}{$u$} & 212 & $1.27\times10^{-3}$ & 3.60e-05 & 5.83e-03 & 2.75e-04 \\
& & 478 & $1.32\times10^{-3}$ & 5.80e-05 & 6.18e-03 & 1.78e-03 \\
\cmidrule(lr){2-7}

& \multirow{2}{*}{$v$} & 212 & $2.59\times10^{-1}$ & 4.96e-03 & 5.82e-01 & 4.54e-02 \\
& & 478 & $2.13\times10^{-1}$ & 9.74e-03 & 5.86e-01 & 1.73e-01 \\
\cmidrule(lr){2-7}

& \multirow{2}{*}{$w$} & 212 & $1.36\times10^{-1}$ & 3.67e-03 & 4.50e-01 & 3.09e-02 \\
& & 478 & $1.77\times10^{-1}$ & 1.05e-02 & 6.63e-01 & 3.02e-01 \\
\bottomrule
\end{tabular}

\caption{Component-wise nMSE assessment for PINN and LVV over full 3D and plane snapshots ($P_s=4$ and $P_s=8$).}
\label{tab:nmse_summary_all}

\end{table*}

\begin{table*}[t]
\centering
\small

\begin{tabular}{@{}llSSSS@{}}
\toprule
\textbf{Domain} & \textbf{Channel} &
\multicolumn{2}{c}{$P_s=4$} &
\multicolumn{2}{c}{$P_s=8$} \\
\cmidrule(lr){3-4} \cmidrule(lr){5-6}

& &
\multicolumn{1}{c}{PINN} &
\multicolumn{1}{c}{LVV} &
\multicolumn{1}{c}{PINN} &
\multicolumn{1}{c}{LVV} \\
\midrule

\multirow{3}{*}{\begin{tabular}[c]{@{}l@{}}Training Snapshots\\(0--424)\end{tabular}}
& $u$ & 1.02e-3 & 4.35e-4 & 4.78e-3 & 7.46e-3 \\
& $v$ & 2.00e-1 & 2.99e-2 & 6.68e-1 & 6.39e-1 \\
& $w$ & 1.43e-1 & 3.28e-2 & 6.35e-1 & 9.36e-1 \\

\midrule

\multirow{3}{*}{\begin{tabular}[c]{@{}l@{}}Validation Snapshots\\(425--530)\end{tabular}}
& $u$ & 1.07e-3 & 4.47e-4 & 4.89e-3 & 6.40e-3 \\
& $v$ & 2.06e-1 & 3.28e-2 & 6.72e-1 & 5.15e-1 \\
& $w$ & 1.45e-1 & 3.53e-2 & 6.36e-1 & 6.45e-1 \\

\bottomrule
\end{tabular}

\caption{Summary of full 3D nMSE across 77 planes for training (first 425) and validation (remaining 106) snapshots.}
\label{tab:nmse_summary_train_val}

\end{table*}

\begin{figure*}[!t]
\centering

\begin{subfigure}{0.9\textwidth}
    \centering
    \includegraphics[height=0.24\textheight,keepaspectratio]{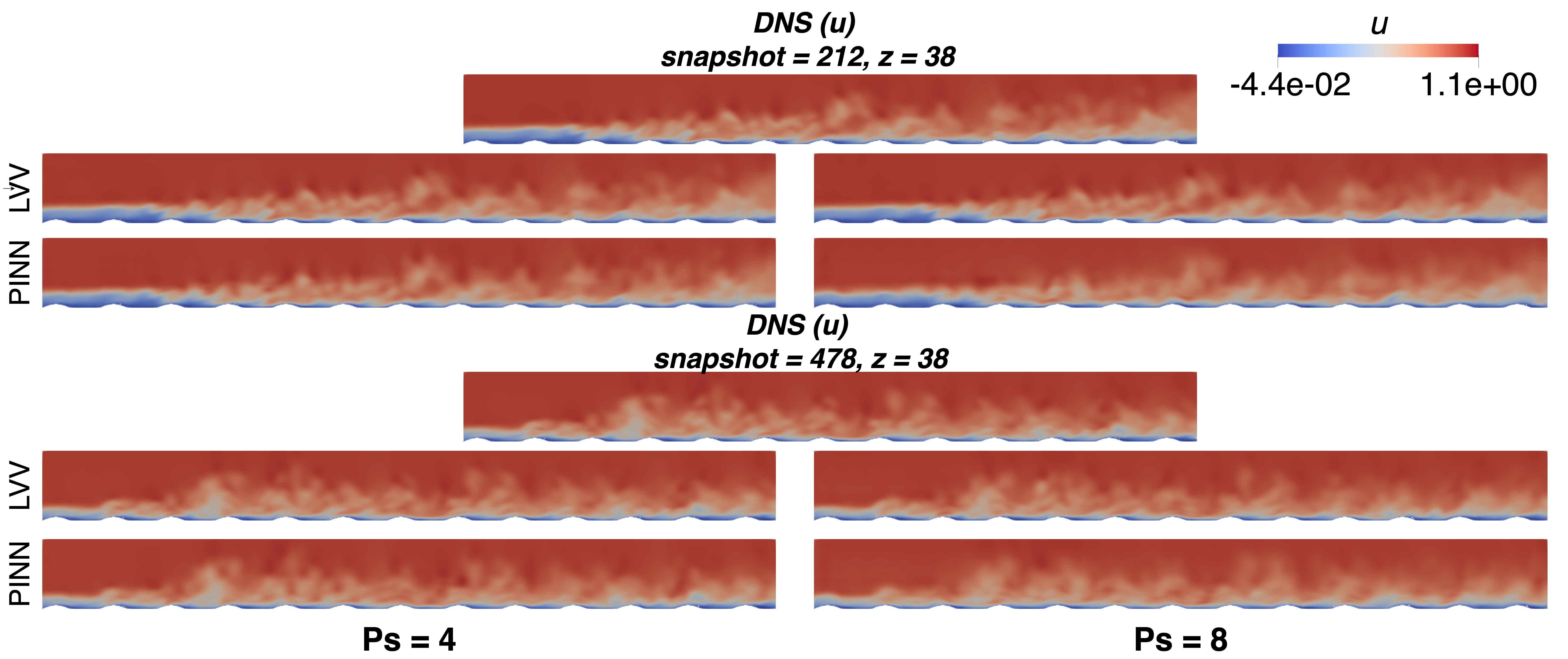}
    \caption{x-component ($u$)}
\end{subfigure}

\vspace{0.2cm}

\begin{subfigure}{0.9\textwidth}
    \centering
    \includegraphics[height=0.24\textheight,keepaspectratio]{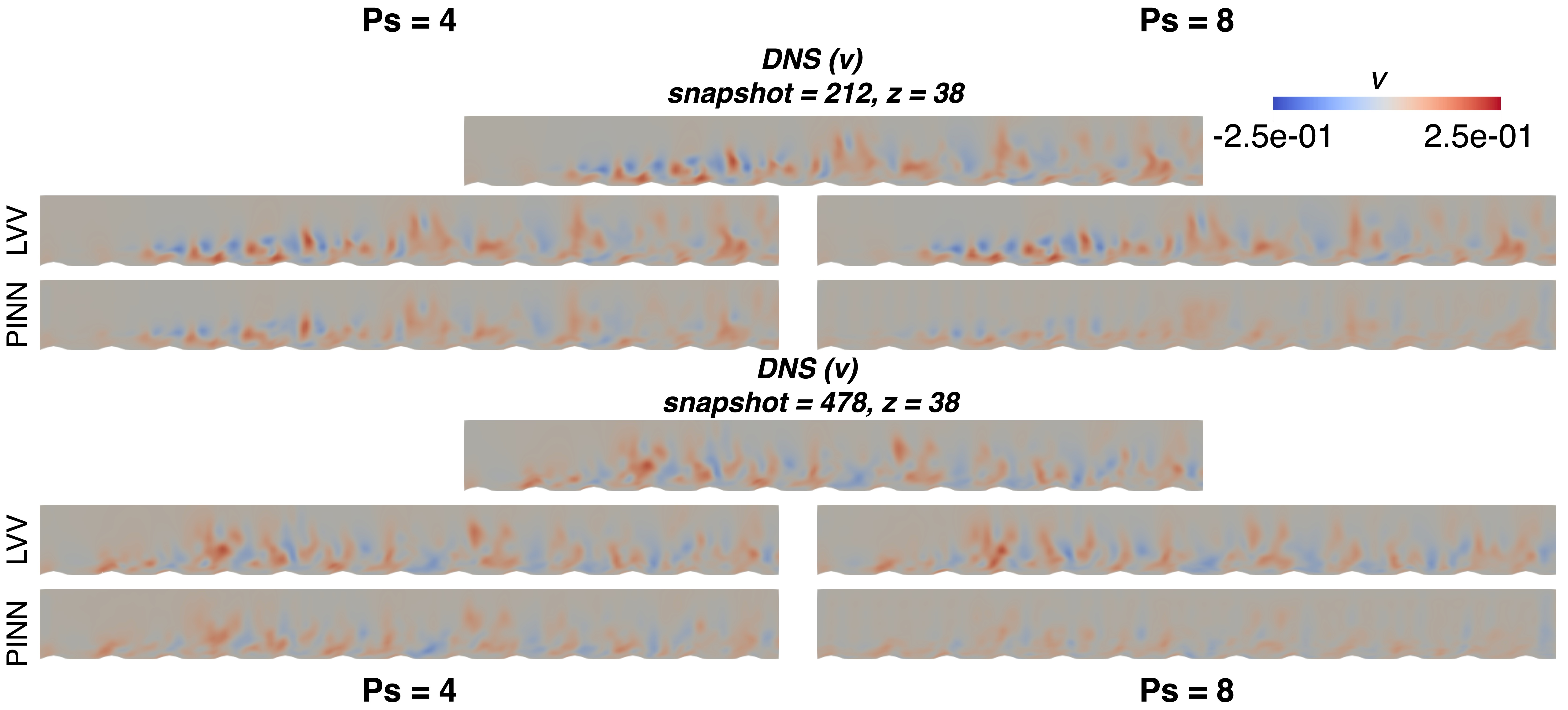}
    \caption{y-component ($v$)}
\end{subfigure}

\vspace{0.2cm}

\begin{subfigure}{0.9\textwidth}
    \centering
    \includegraphics[height=0.24\textheight,keepaspectratio]{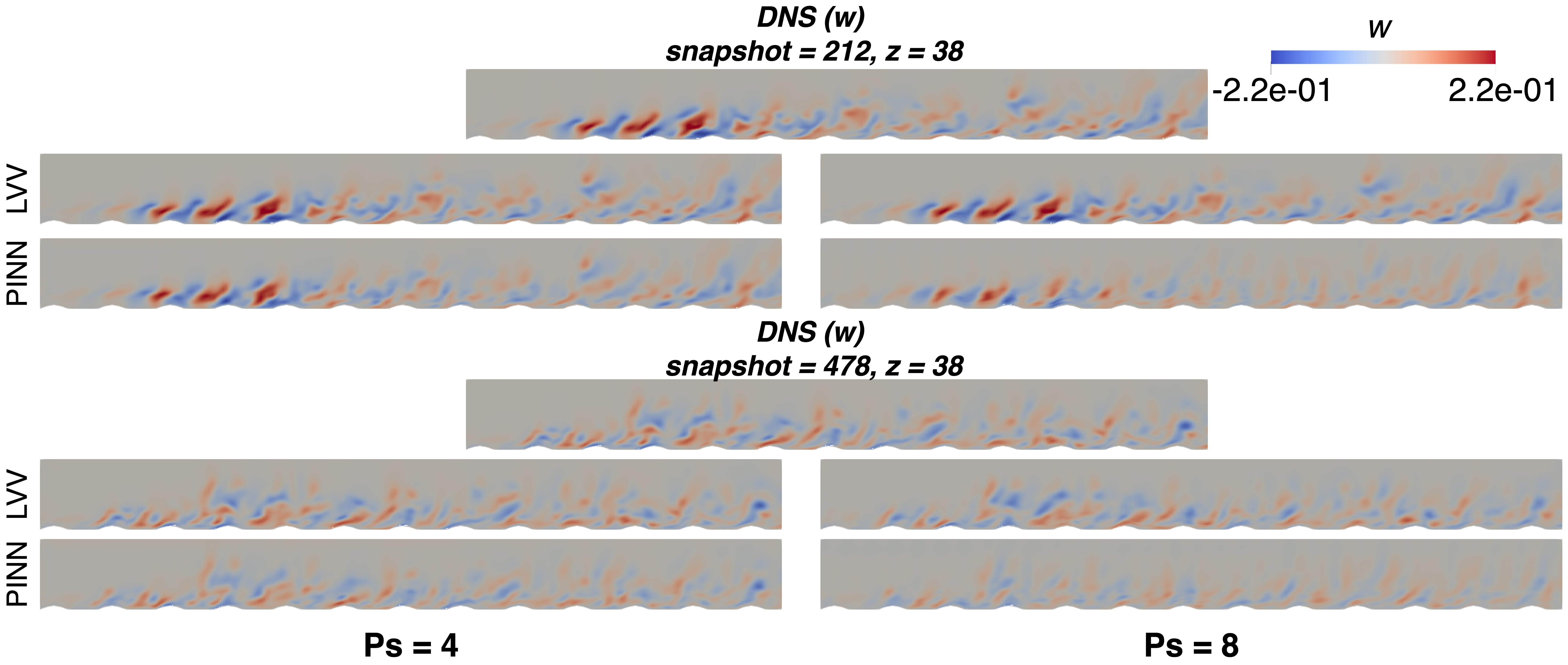}
    \caption{z-component ($w$)}
\end{subfigure}

\caption{Velocity components along $z=38$ (plane within the training grid) comparing LVV and PINN against DNS at snapshots 212 (training) and 478 (validation). Panels (a), (b), and (c) correspond to the $u$, $v$, and $w$ components, respectively.}

\label{fig:velocity_lvv_pinn_combined_z_38}

\end{figure*}

Figure~\ref{fig:mse_line_plot} (a) and (b) presents the snapshot-wise MSE for the training $z$-slices ($33$--$43$), outlined in green in Figure~\ref{fig:non_uniform_grid}, and the validation $z$-slices ($0$--$32 \cup 44$--$76$), which lie outside the training grid, for both upscaling factors $P_s=4$ and $8$, respectively. Figure ~\ref{fig:mse_line_plot} (c) and (d) represent the spanwise MSE along $z$, averaged across training (0 - 424) and validation (425 - 530) snapshots, respectively. The training domain is shaded green, and the validation domain is shaded red. At $P_s=4$, LVV maintains low error across most of the domain, indicating good reconstruction performance both within and beyond the training region. However, at $P_s=8$, the performance of LVV deteriorates outside the training domain, as indicated by the increase in MSE over the validation snapshots in Figure~\ref{fig:mse_line_plot} (a) and the larger error over the validation $z$-planes in Figure~\ref{fig:mse_line_plot} (b). From figure ~\ref{fig:mse_line_plot} (c) and (d), the error for LVV drastically increases as it approaches the right boundary at $P_s = 8$. In contrast, PINN maintains a much more stable error trend across the spanwise $z$ direction and also across the validation snapshots. This suggests that the held-out extrapolation robustness of LVV becomes more limited as sparsity increases. In contrast, PINN does not exhibit such a pronounced mismatch between the training and validation regions. For both $P_s=4$ and $8$, PINN shows a more uniform error distribution across the training and validation $z$-slices and snapshots. Moreover, at $P_s=8$, the overall MSE of PINN over the validation $z$-planes appears lower than that of LVV, as seen in Figure~\ref{fig:mse_line_plot} (b). This suggests that PINN exhibits more stable out-of-training-domain behaviour under increased sparsity, even though its absolute reconstruction accuracy remains lower for some velocity components, as discussed below. 

While Figure~\ref{fig:mse_line_plot} provides a spatial and temporal view of error trends through MSE, Tables~\ref{tab:nmse_summary_all} and~\ref{tab:nmse_summary_train_val} quantify component-wise reconstruction accuracy using the normalised mean squared error metric (nMSE) described in equation \ref{eq:nmse}.  For each velocity component $\phi\in\{u,v,w\}$, nMSE over the $x$--$y$--$z$ domain is defined as
\begin{equation}
\label{eq:nmse}
\mathrm{nMSE}_{\phi}(s)=
\frac{
\sum_{i=1}^{N_{xyz}}
\left(
\phi_{\mathrm{pred}}^{(i)}(s)
-
\phi_{\mathrm{DNS}}^{(i)}(s)
\right)^2
}
{
\sum_{i=1}^{N_{xyz}}
\left(
\phi_{\mathrm{DNS}}^{(i)}(s)
\right)^2+\epsilon
},
\end{equation}
where $N_{xyz}$ denotes the number of spatial grid points and $\epsilon$ is a small constant introduced to avoid division by zero. This normalisation accounts for differences in the magnitude of the reference field, thereby providing a fairer measure of reconstruction quality across velocity components. To quantify reconstruction quality, we report component-wise errors at $z=38$ and $z=19$, as well as across all 77 spanwise planes. Table~\ref{tab:nmse_summary_all} reports the nMSE values, computed using Eq.~\ref{eq:nmse}, for two representative snapshots: $s=212$ from the training snapshots and $s=478$ from the validation snapshots. Table~\ref{tab:nmse_summary_train_val} summarises the corresponding nMSE values across all 77 spanwise planes averaged over both training and validation snapshots. Key takeaways can be summarized as follows:
\begin{figure*}[!t]
\centering

\begin{subfigure}{0.9\textwidth}
    \centering
    \includegraphics[height=0.24\textheight,keepaspectratio]{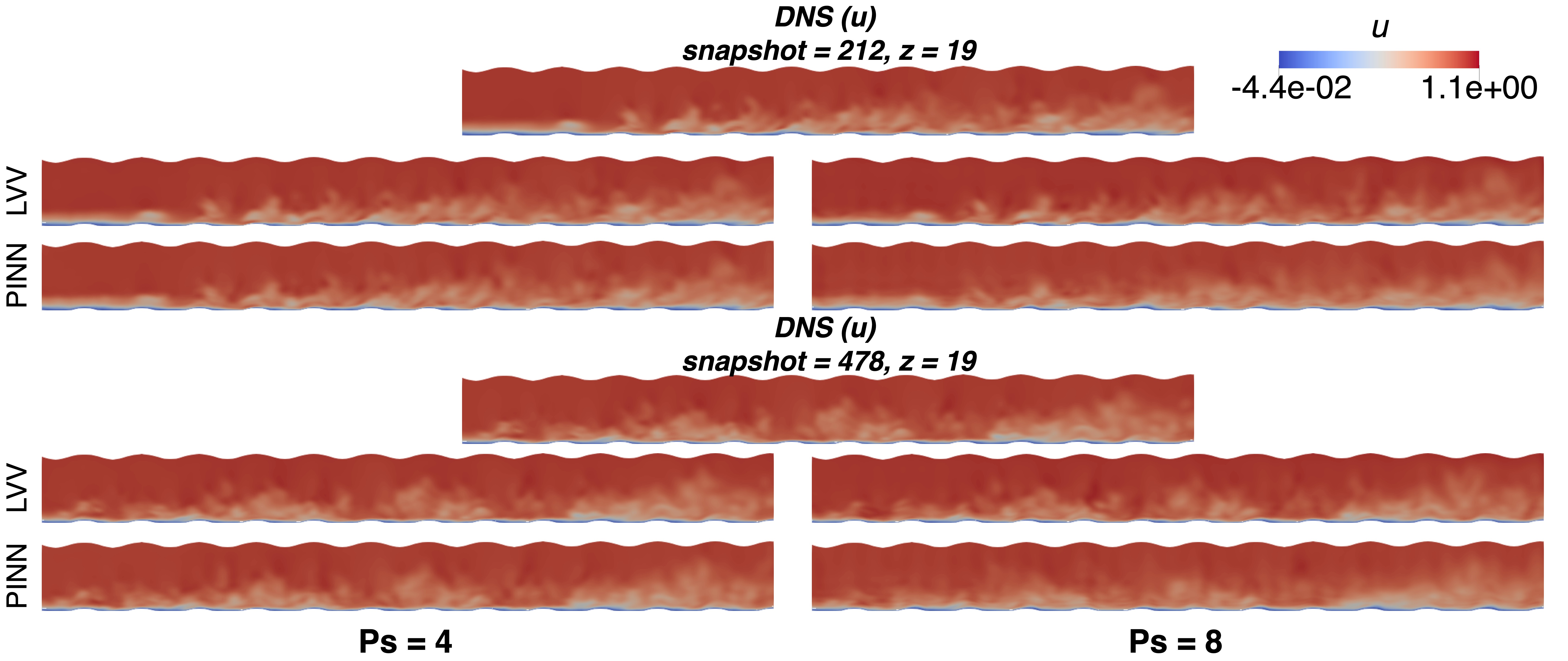}
    \caption{x-component ($u$)}
\end{subfigure}

\vspace{0.2cm}

\begin{subfigure}{0.9\textwidth}
    \centering
    \includegraphics[height=0.24\textheight,keepaspectratio]{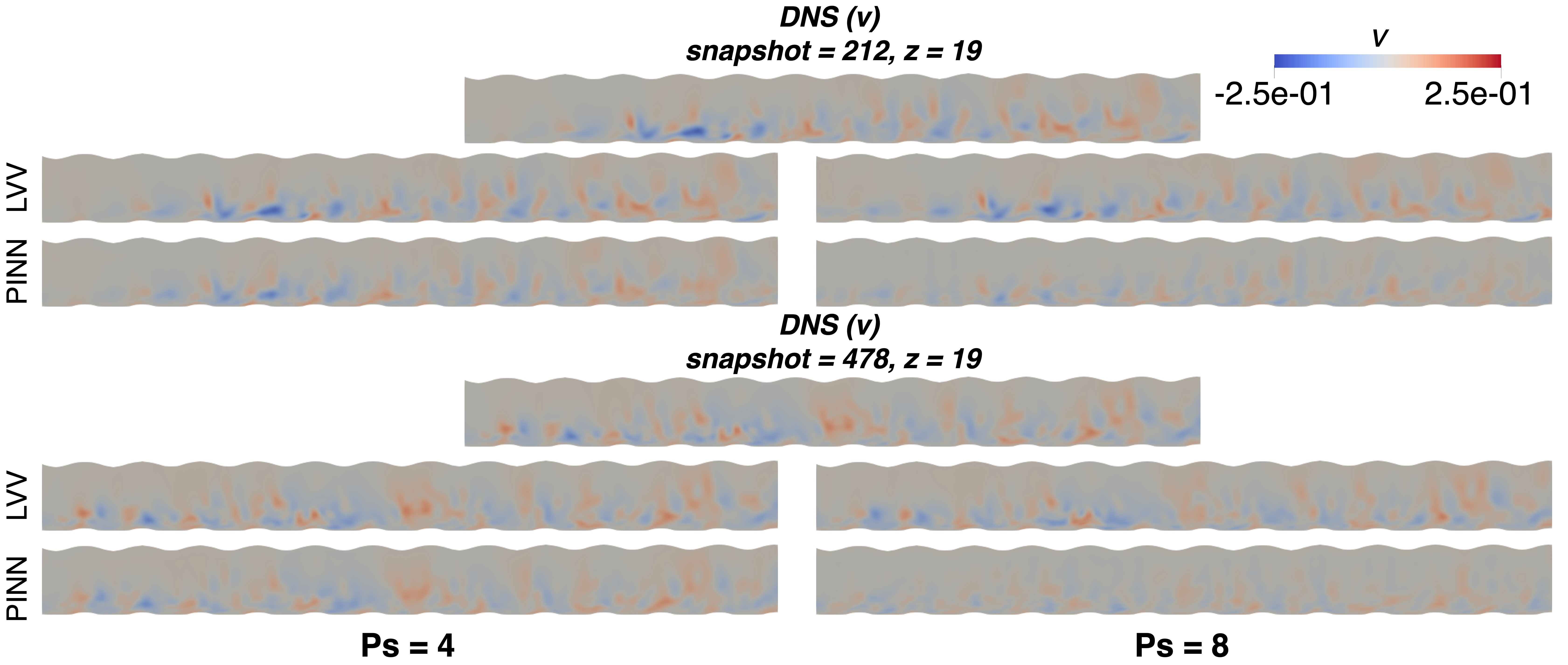}
    \caption{y-component ($v$)}
\end{subfigure}

\vspace{0.2cm}

\begin{subfigure}{0.9\textwidth}
    \centering
    \includegraphics[height=0.24\textheight,keepaspectratio]{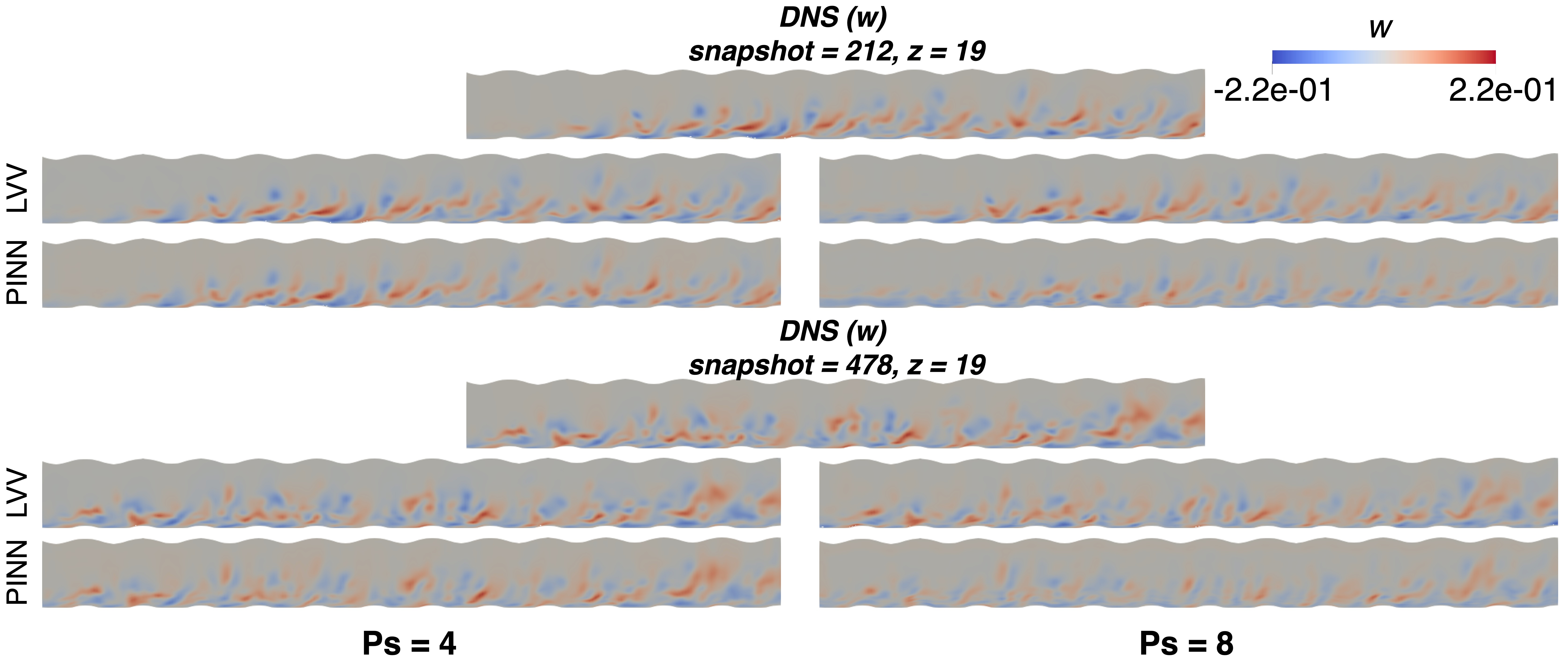}
    \caption{z-component ($w$)}
\end{subfigure}

\caption{Velocity components along $z=19$ (plane outside the training grid) comparing LVV and PINN against DNS at snapshots 212 (training) and 478 (validation). Panels (a), (b), and (c) correspond to the $u$, $v$, and $w$ components, respectively.}

\label{fig:velocity_lvv_pinn_combined_z_19}

\end{figure*}

\begin{enumerate}
    \item For both $P_s=4$ and $8$, the nMSE of the streamwise velocity component ($u$) is consistently lower than that of the transverse components ($v$ and $w$) at both snapshots $s=212$ and $s=478$ (Table ~\ref{tab:nmse_summary_all}) as well as across the entire domain for both training and validation snapshots (Table ~\ref{tab:nmse_summary_train_val}) for both models.
    \item At $ P_s = 4$, the difference between the streamwise ($u$) and traverse components ($v,w$) is more pronounced for PINN than for LVV, indicating a stronger degradation in reconstructing $v$ and $w$ in the absence of high-resolution labels. For PINN, the reconstruction errors in $v$ and $w$ are already substantial at $P_s=4$ and increase further at $P_s=8$. This behaviour is consistent with the lower amplitudes and stronger intermittency of the transverse velocity components in transitional and turbulent boundary-layer flows. Similar trends have also been reported in previous reconstruction studies~\cite{yousif2023deep}. 
    \item At $P_s=8$, the drop in performance for LVV may indicate reduced extrapolation robustness due to overfitting; for PINN, however, the results suggest that PINN has greater difficulty in recovering the velocity components under sparse sampling conditions. Overall, the results indicate that the dominant streamwise component is reconstructed more accurately than the weaker transverse components, and the performance of models decreases as sparsity increases. 
\end{enumerate}

It is important to note that the nMSE values in Table ~\ref{tab:nmse_summary_all} are consistently lower for LVV than for PINN across $z = 19$, $38$ at $P_s = 8$. However, this observation changes when we examine the results across all 77 planes in Table ~\ref{tab:nmse_summary_all} and across both training and validation snapshots in Table ~\ref{tab:nmse_summary_train_val}. This can be summarised briefly by examining the line plots in Figure ~\ref{fig:mse_line_plot} (c) and (d), where, unlike PINN, which shows a more stable error distribution across $z$, LVV error tends to increase as it moves farther from the training grid. This contributes to a higher MSE and nMSE, when comparing across the entire domain. Since $z = 19$ is outside this notoriously high error region, the values vary when comparing across the single $z = 19$, $z = 38$ and across 77 planes.

We further examine instantaneous reconstructions from LVV and PINN in Figures~\ref{fig:velocity_lvv_pinn_combined_z_38} and~\ref{fig:velocity_lvv_pinn_combined_z_19}, which show the three velocity components $(u,v,w)$ on the spanwise planes $z=38$ (within the training grid) and $z=19$ (outside the training grid), at snapshots $s=212$ and $s=478$. Across both snapshots and both planes, both models recover the dominant large-scale structures. In particular, the streamwise velocity component $u$ reconstructed by PINN remains visually close to the DNS reference field, despite the absence of high-resolution velocity supervision. Deviations become more apparent as sparsity increases, especially for the transverse velocity components, which is consistent with the nMSE trends discussed above.

\subsection{Energy spectra\label{subsec:energy_spectra}}
\begin{figure*}[t]
\centering

\begin{subfigure}[t]{0.48\textwidth}
    \centering
    \includegraphics[width=\linewidth]{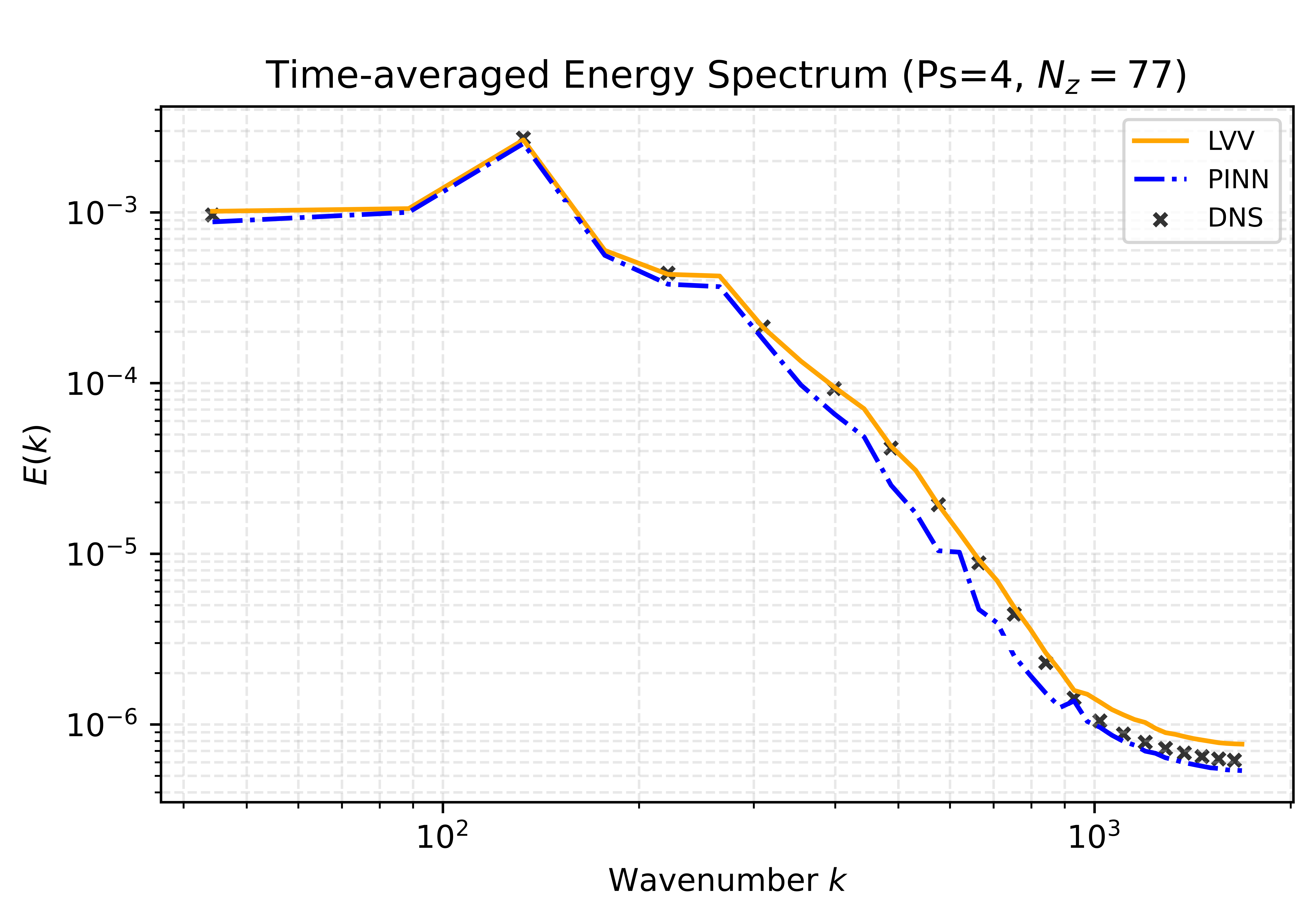}
    \caption{$P_s=4$}
\end{subfigure}
\hfill
\begin{subfigure}[t]{0.48\textwidth}
    \centering
    \includegraphics[width=\linewidth]{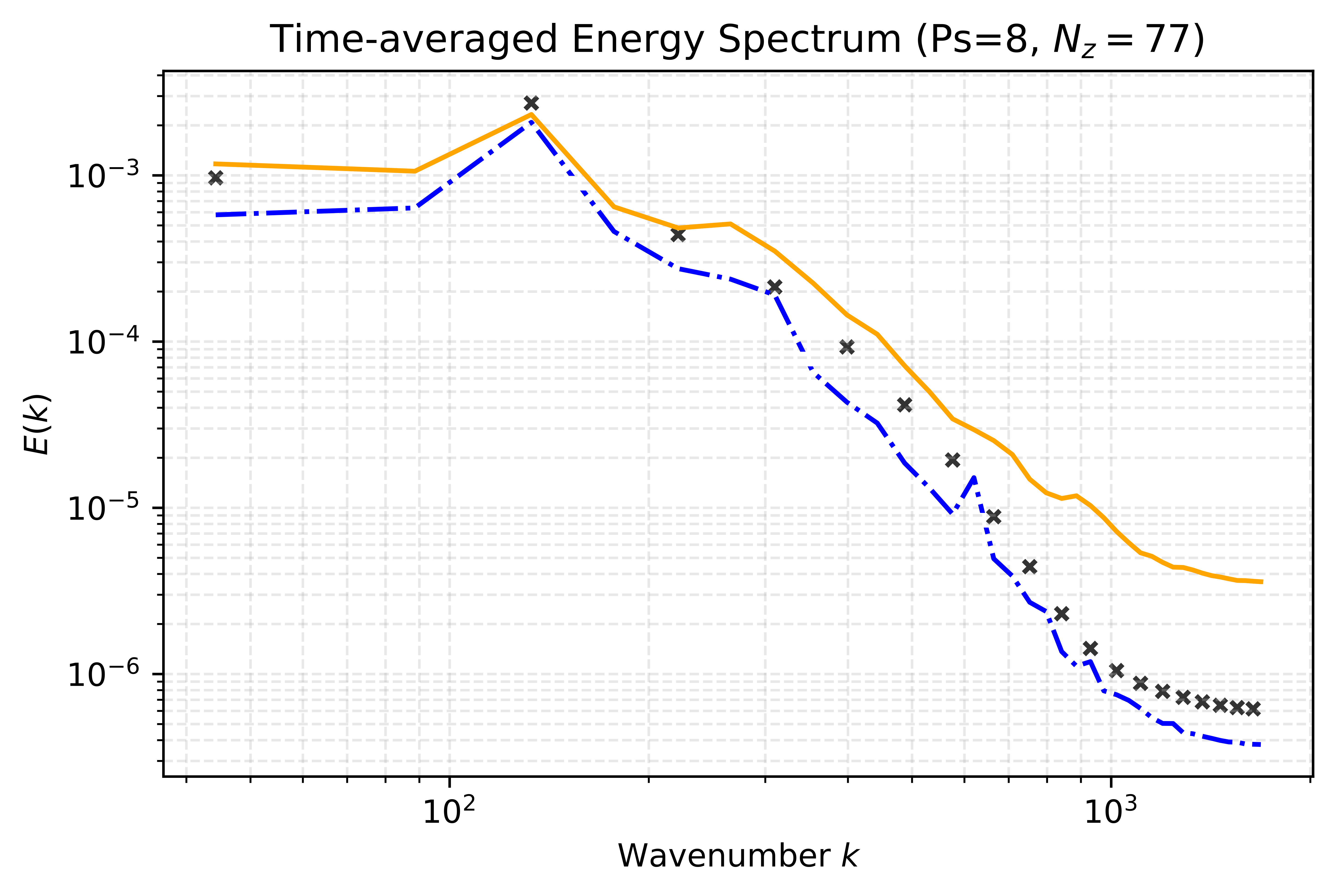}
    \caption{$P_s=8$}
\end{subfigure}

\caption{Spanwise energy spectra for DNS, LVV, and PINN. The spectra illustrate the distribution of turbulent kinetic energy across wavenumbers and are used to assess the extent to which large- and small-scale flow structures are recovered by the reconstruction models as compression increases.}
\label{fig:energy_spectra}
\end{figure*}

Turbulent flows contain a broad range of dynamically relevant scales, and an important measure of reconstruction quality is how well these scales are recovered. To assess this, we compute time-averaged spanwise energy spectra along the periodic $z$ direction over the full domain, \(531\times256\times64\times77\). Figure~\ref{fig:energy_spectra} assesses the time-averaged energy spectra of the reconstructed fields against the DNS reference for LVV (orange) and PINN (blue) at upscaling factors $P_s=4$ and $8$.

As shown in Figure~\ref{fig:energy_spectra} (a), at $P_s=4$, both LVV and PINN models reproduce the large-scale, energy-containing structures well, as indicated by their close agreement with the DNS at low wavenumbers. As the wavenumber increases, deviations become more apparent, particularly for PINN, which exhibits a faster decay of spectral energy at higher $k$, indicating reduced small-scale spectral content. In contrast, LVV maintains closer agreement with the DNS over a wider range of wavenumbers, suggesting better recovery of intermediate and smaller-scale structures at the lower compression level.

As shown in Figure~\ref{fig:energy_spectra} (b), at $P_s=8$, spectral degradation is observed for both models, consistent with increased sparsity. While both methods still capture the large-scale trend reasonably well, discrepancies become more evident at intermediate and high wavenumbers. Here, the two models exhibit different error characteristics. LVV tends to retain comparatively higher energy at larger wavenumbers, with a slower decay relative to DNS, whereas PINN shows a more rapid loss of spectral energy and reduced high-wavenumber content. Thus, both models deviate from DNS at higher compression, but in different ways: LVV tends to over-retain high-wavenumber energy, whereas PINN exhibits reduced small-scale spectral content. The two approaches exhibit different error mechanisms rather than a simple ranking in performance.

Overall, these results indicate that spectral fidelity decreases with increasing compression, with the degradation becoming most evident at smaller scales. Large-scale turbulent structures remain comparatively well reconstructed by both approaches, whereas the recovery of fine-scale motions remains more challenging. These spectral trends are consistent with the reconstruction errors and held-out extrapolation behaviour discussed in Section~\ref{subsec:inst_vel}.

\subsection{Turbulent statistics}
Beyond point-wise errors and spectral fidelity, it is important to assess whether the reconstructed fields recover key turbulence statistics and coherent structures. To this end, three complementary diagnostics are examined. First, turbulent kinetic energy is used to assess the recovery of the integral energy content of the flow. Second, Reynolds stresses are analysed to evaluate whether the reconstructed fields retain turbulent momentum fluctuations and anisotropic statistical structure. Third, three-dimensional $Q$-criterion iso-surfaces are examined to assess recovery of coherent vortical structures. Together, these diagnostics provide a progressively richer assessment of reconstruction fidelity, from integral energy measures to statistical structure and instantaneous flow topology.

\subsubsection{Maximum Turbulent Kinetic Energy\label{subsubsec:tke}}

Turbulent kinetic energy is defined as

\begin{equation}
k= \frac{1}{2}\left(\overline{u'^2}+\overline{v'^2}+\overline{w'^2}\right),
\label{eq:tke}
\end{equation}
where $u' = u-\overline{u}$, $v' = v-\overline{v}$, and $w' = w-\overline{w}$ are the fluctuating velocity components, and $\overline{u}$, $\overline{v}$, and $\overline{w}$ denote the corresponding mean velocity components.

\begin{figure*}[!t]
\centering

\begin{subfigure}{0.9\textwidth}
    \centering
    \includegraphics[width=0.9\textwidth]{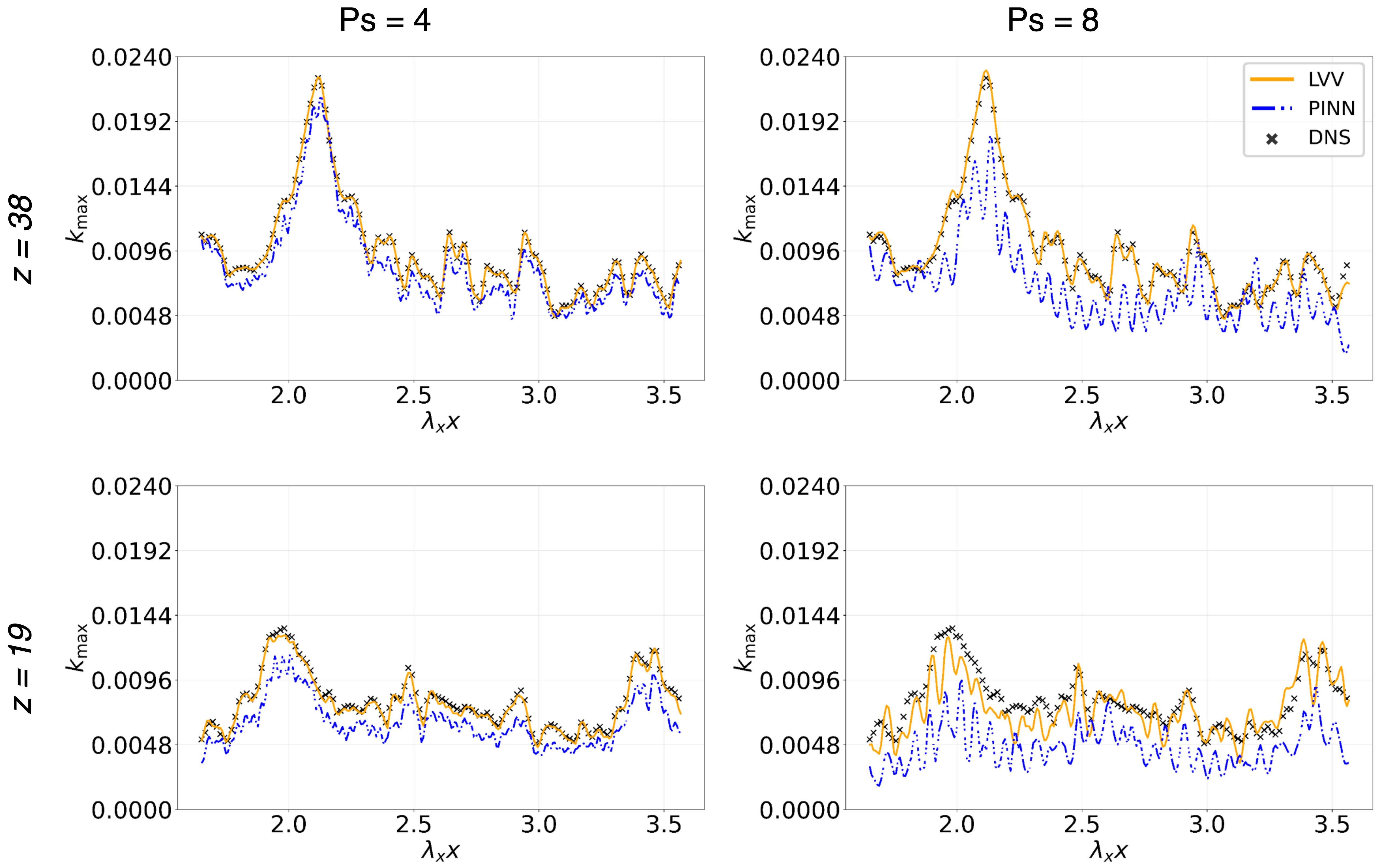}
    \caption{Training snapshots}
\end{subfigure}

\vspace{0.2cm}

\begin{subfigure}{0.9\textwidth}
    \centering
    \includegraphics[width=0.9\textwidth]{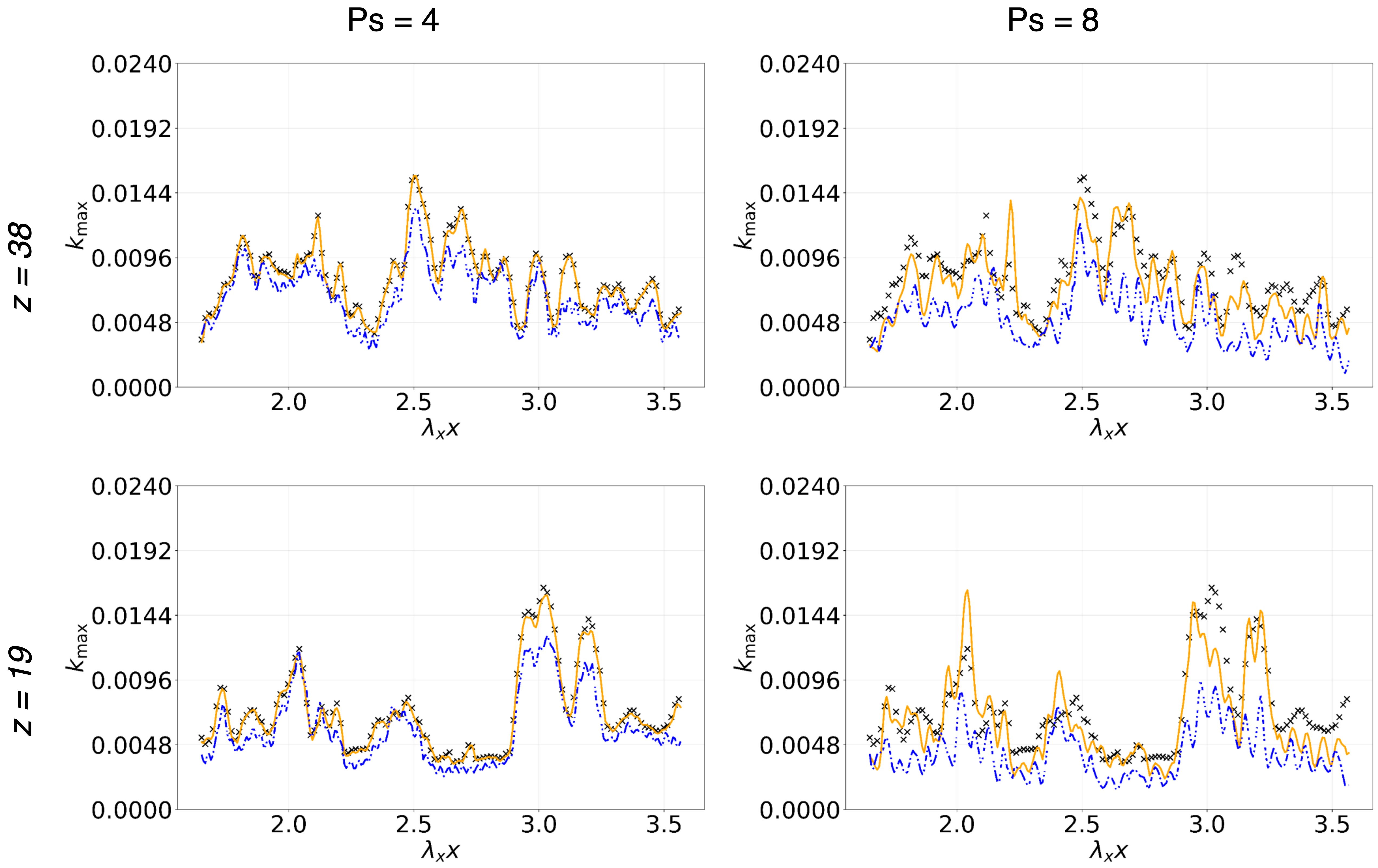}
    \caption{Validation snapshots}
\end{subfigure}

\caption{Maximum turbulent kinetic energy ($k_{\max}$) profiles at spanwise planes $z=38$ and $z=19$ for DNS, LVV, and PINN at compression levels $P_s=4$ and $P_s=8$. Panels (a) and (b) correspond to training and validation snapshots, respectively.}

\label{fig:tke}

\end{figure*}

Figure~\ref{fig:tke} presents the evolution of the streamwise variation of maximum turbulent kinetic energy ($k_{\max}$) at $z=38$ and $z=19$ for both training and validation snapshots. At $P_s=4$, LVV recovers $k_{\max}$ reasonably well at both planes for both training and validation snapshots, while PINN follows the overall trend but tends to underestimate the peak values. At $P_s=8$, deviations increase for both models, particularly at $z=19$ and in the validation snapshots, indicating reduced fidelity to conserve $k_{max}$ under stronger compression.

A notable feature is the systematic underestimation of $k_{\max}$ by PINN, which is consistent with the larger reconstruction errors observed earlier for the transverse velocity components $v$ and $w$ in Section~\ref{subsec:inst_vel}. Since turbulent kinetic energy depends on contributions from all three fluctuating components, errors in reconstructing $v$ and $w$ directly affect the predicted $k_{\max}$. Overall, the TKE results indicate that integral turbulence statistics are reasonably recovered at lower compression, whereas both models exhibit increasing deviations as compression increases.

\subsubsection{Reynolds stresses{\label{subsec:reynolds_stresses}}}

The Reynolds stresses quantify turbulent momentum fluctuations and provide a more stringent statistical diagnostic than integral energy measures alone. The Reynolds stress tensor is defined as

\begin{equation}
\label{eq:reynolds_stress_matrix}
\mathbf{R}
=
-\rho
\begin{bmatrix}
\overline{u'u'} & \overline{u'v'} & \overline{u'w'}\\
\overline{v'u'} & \overline{v'v'} & \overline{v'w'}\\
\overline{w'u'} & \overline{w'v'} & \overline{w'w'}
\end{bmatrix}
\end{equation}

\begin{figure*}[!t]
\centering

\begin{subfigure}{0.9\textwidth}
    \centering
    \includegraphics[width=0.9\textwidth]{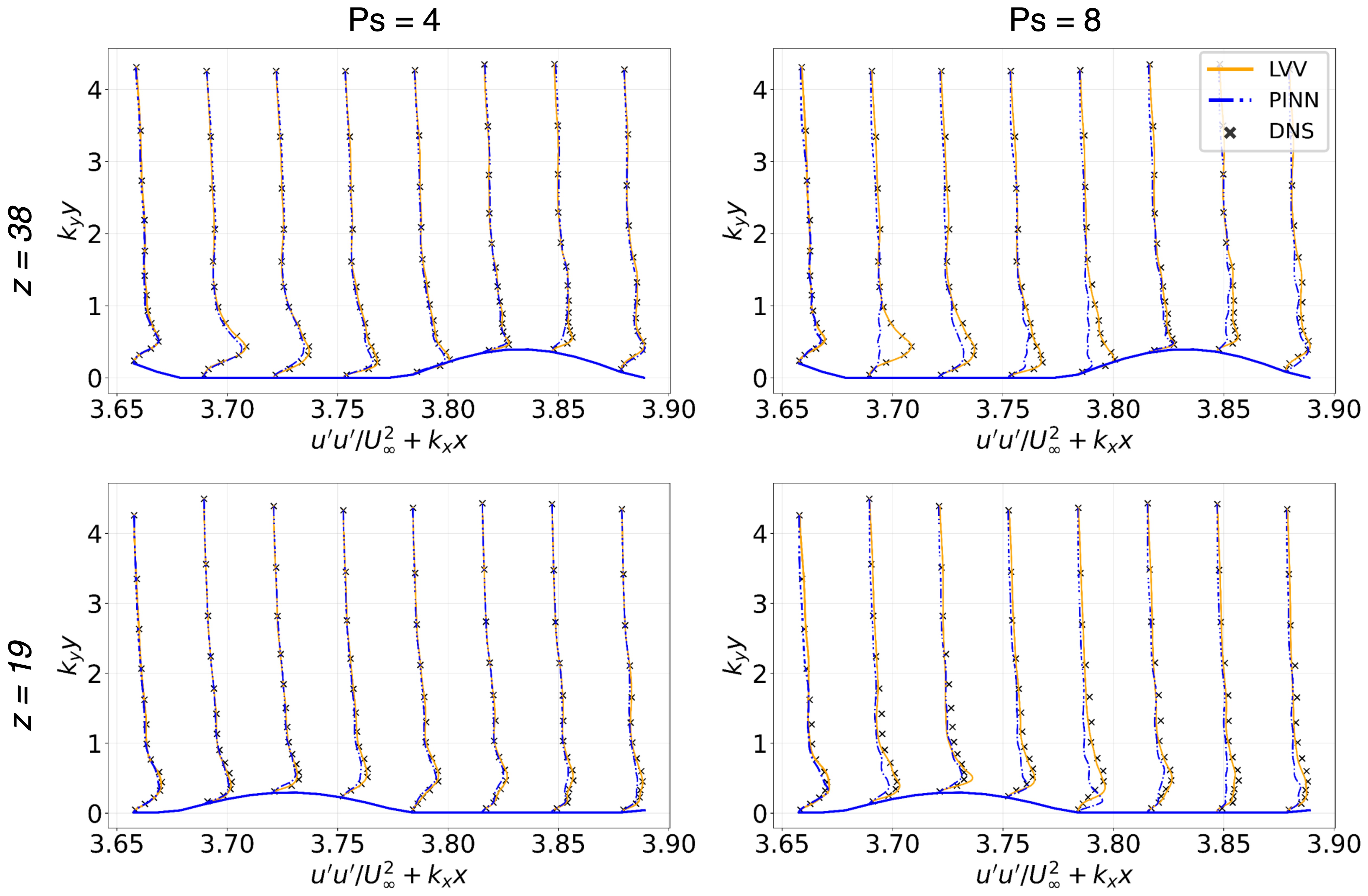}
    \caption{Training snapshots}
\end{subfigure}

\vspace{0.2cm}

\begin{subfigure}{0.9\textwidth}
    \centering
    \includegraphics[width=0.9\textwidth]{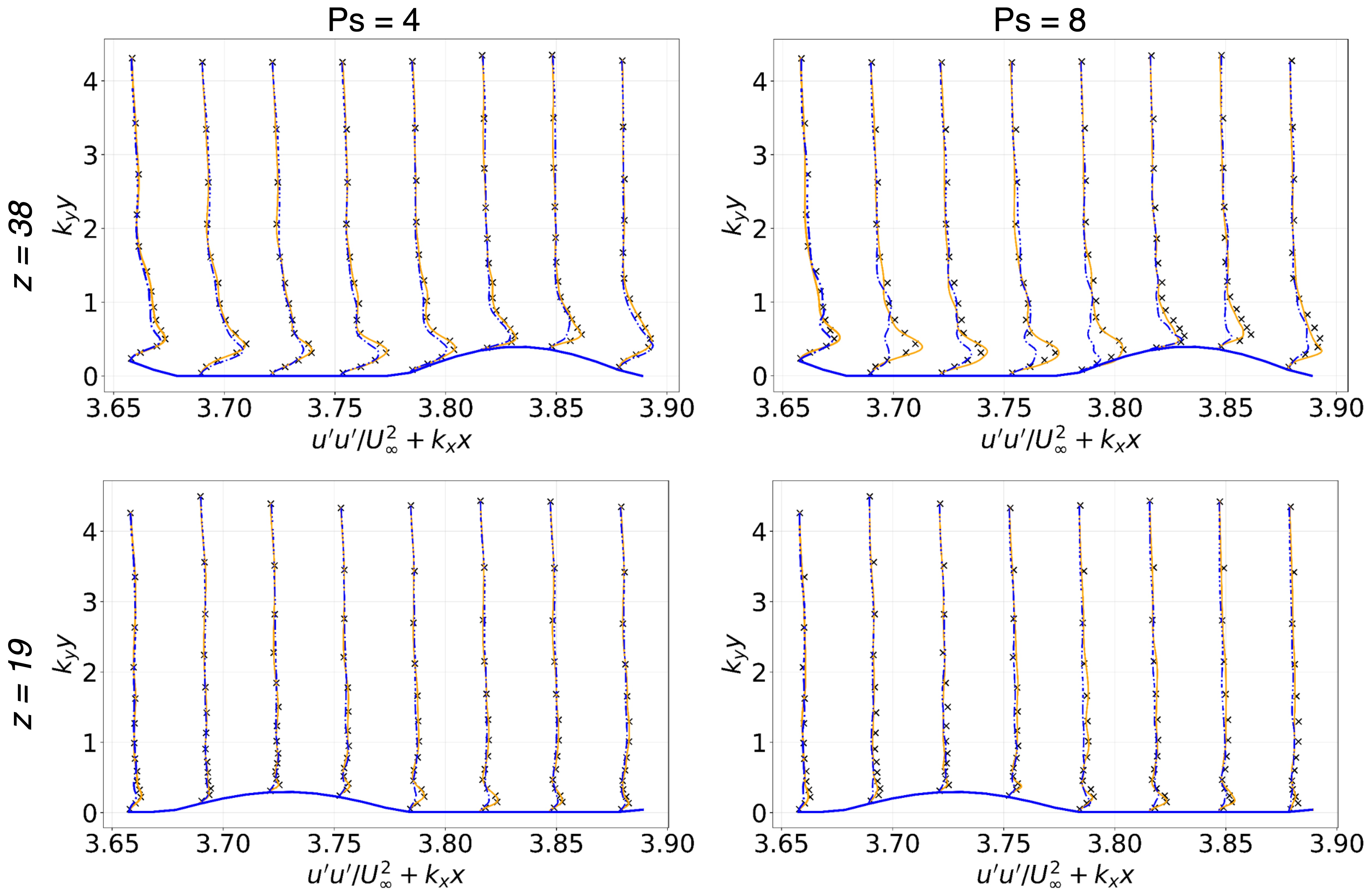}
    \caption{Validation snapshots}
\end{subfigure}

\caption{Carpet plots of the Reynolds stress component $u'u'$ at spanwise planes $z=38$ (within the training grid) and $z=19$ (outside the training grid) for DNS, LVV, and PINN at compression levels $P_s=4$ and $P_s=8$. Panels (a) and (b) correspond to training and validation snapshots, respectively.}

\label{fig:reynolds_stresses}

\end{figure*}

where $u'$, $v'$, and $w'$ denote the streamwise, wall-normal, and spanwise velocity fluctuations, respectively. Figure~\ref{fig:reynolds_stresses} presents carpet plots of the Reynolds stress component $u'u'$ over a roughness element at $z=38$ and $z=19$ for training and validation snapshots.

At $P_s=4$, both the LVV and PINN models reproduce the Reynolds-stress profiles reasonably well, with LVV generally showing closer agreement with DNS, particularly near the near-wall peak regions. This indicates that the dominant Reynolds-stress structure is recovered at the lower compression level for both the models.

At $P_s=8$, deviations become more evident for both models, with LVV showing better agreement than for PINN. PINN exhibits increased deviations, especially near the near-wall peak. Overall, Reynolds-stress fidelity deteriorates with increasing compression, with LVV generally retaining better agreement at lower compression, while both models show limitations at $P_s=8$.

These observations are broadly consistent with the TKE and spectral trends, suggesting that while integral and second-order turbulence statistics remain reasonably well recovered at lower compression, their fidelity degrades progressively as sparsity increases.

\subsubsection{Coherent structures via Q-criterion iso-surfaces\label{subsec:Qisosurface_generalization}}

To further assess the recovery of instantaneous flow topology, three-dimensional iso-surfaces of the $Q$-criterion are examined. The $Q$-criterion is defined as
\begin{equation}
Q=\frac{1}{2}\left(\|\boldsymbol{\Omega}\|^2-\|\mathbf{S}\|^2\right),
\end{equation}
where $\boldsymbol{\Omega}$ and $\mathbf{S}$ denote the antisymmetric
(rotation-rate) and symmetric (strain-rate) parts of the velocity-gradient
tensor, respectively. Positive values of $Q$ identify regions in which local
rotation dominates over strain.

\begin{figure*}[t]
\centering

\begin{subfigure}{0.9\textwidth}
    \centering
    \includegraphics[width=\textwidth]{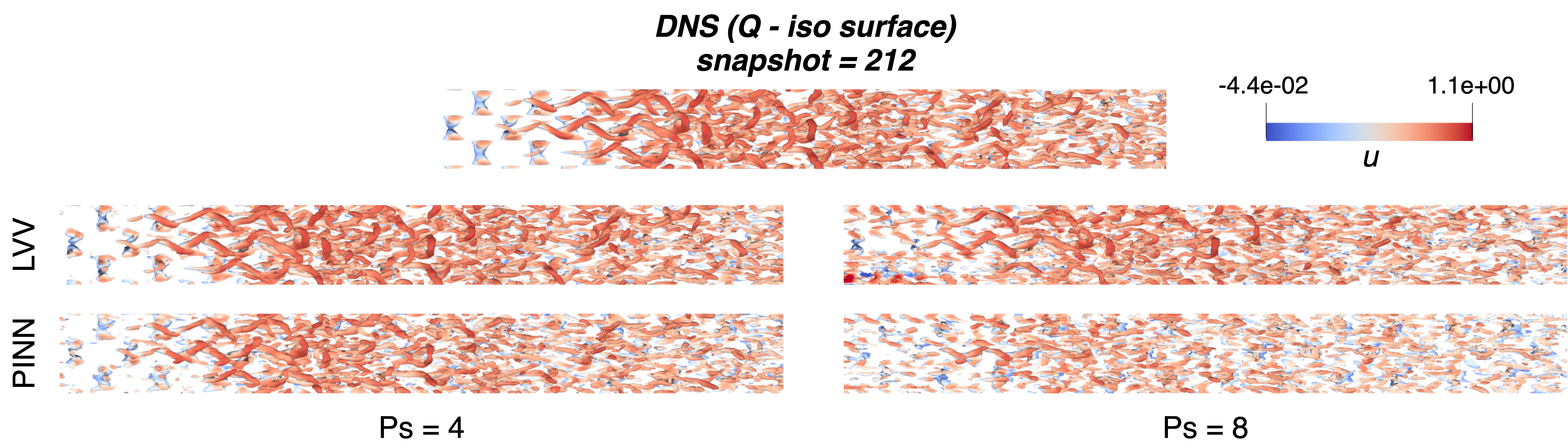}
    \caption{Snapshot $s=212$.}
\end{subfigure}

\vspace{0.5cm}

\begin{subfigure}{0.9\textwidth}
    \centering
    \includegraphics[width=\textwidth]{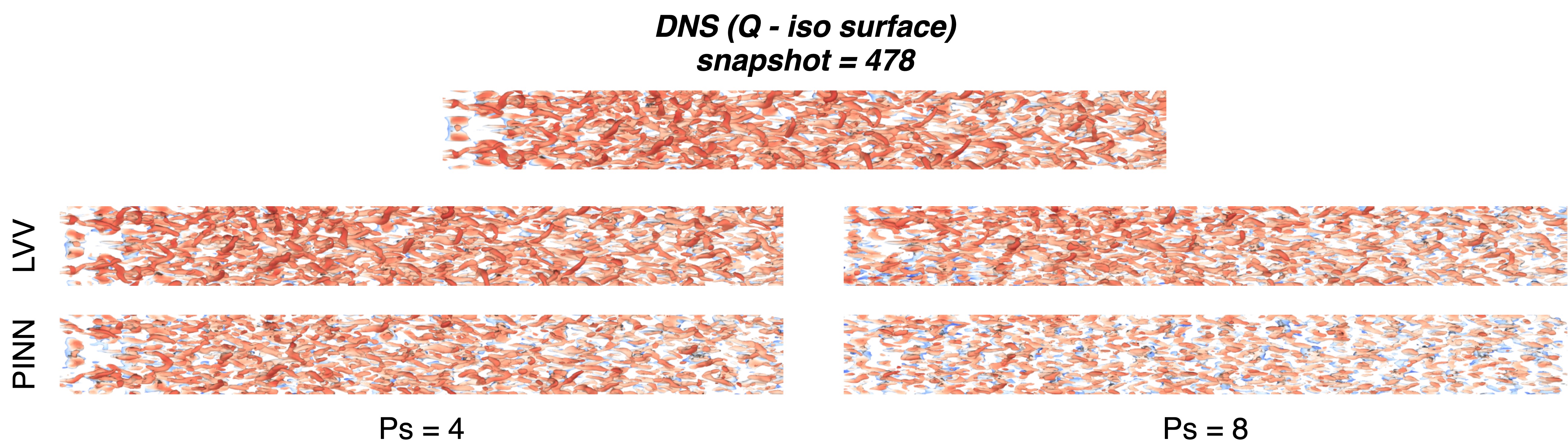}
    \caption{Snapshot $s=478$.}
\end{subfigure}

\caption{Full-domain comparison of $Q$-criterion iso-surfaces ($Q=100$) for
the DNS reference, LVV, and PINN at representative snapshots $s=212$ and
$s=478$. Results are shown for compression levels $P_s=4$ and $P_s=8$.
Iso-surfaces are coloured by the streamwise velocity component $u$ using an
identical colour scale for the DNS and reconstructed fields.}
\label{fig:q_iso_surfaces_lvv_pinn}
\end{figure*}

Figure~\ref{fig:q_iso_surfaces_lvv_pinn} compares the full-domain
$Q$-criterion iso-surfaces reconstructed by LVV and PINN with the DNS
reference at snapshots $s=212$ and $s=478$. At $P_s=4$, both reconstruction
methods recover a substantial portion of the rotation-dominated regions,
although LVV preserves the spatial continuity and organisation of the
vortical field more closely than PINN. PINN captures several coherent
structures but exhibits stronger local fragmentation and displacement.

At $P_s=8$, the degradation of the reconstructed vortical field becomes more
pronounced. LVV retains portions of the larger coherent structures, although
their continuity is reduced relative to DNS. PINN shows a stronger loss of
connected vortical features and a more fragmented topology. The full-domain
comparison therefore indicates that the recovery of fine-scale vortical
structures becomes increasingly difficult as the compression level increases.

\begin{figure*}[t]
\centering 
\includegraphics[width=\textwidth]{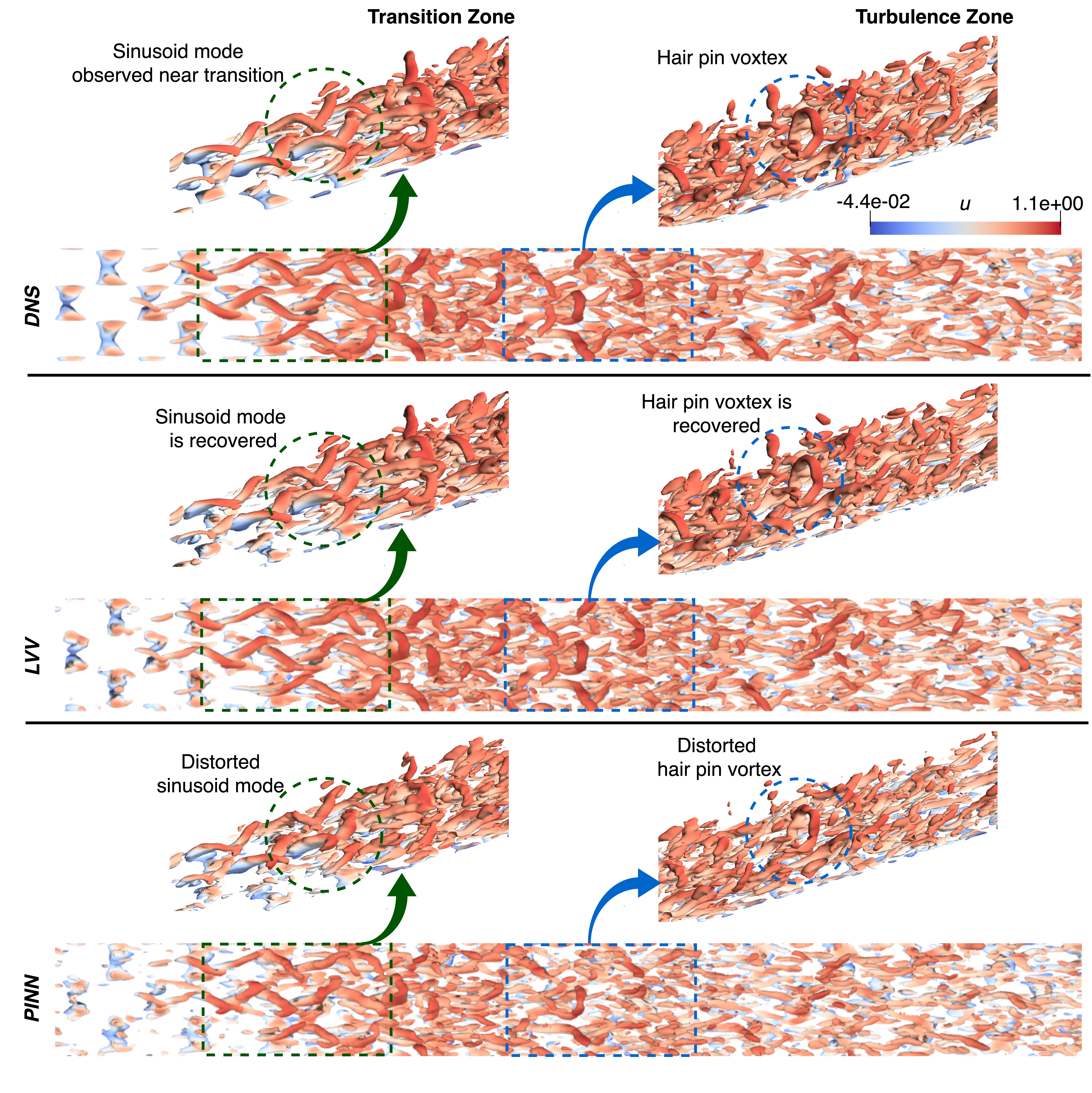}
\caption{Zoomed comparison of $Q$-criterion iso-surfaces at snapshot
$s=212$ for $P_s=4$. The selected subdomains correspond to a
near-transition region, where the DNS field exhibits a developing sinusoid
vortical structure, and a downstream turbulent region containing a
representative hairpin-like vortex. The identical physical subdomains are
extracted from DNS, LVV, and PINN to facilitate a direct comparison of
local coherent-structure recovery. The near-transition box spans
$(x_{\mathrm{start}},x_{\mathrm{end}})=(1.3,1.5)$, while the downstream
turbulent box spans $(x_{\mathrm{start}},x_{\mathrm{end}})=(1.7,1.9)$.
All panels use the same iso-surface threshold ($Q=100$), viewing direction,
and streamwise-velocity colour scale.}
\label{fig:ps_4_q_zoom_s212}
\end{figure*}

\begin{figure*}[t]
\centering
\includegraphics[width=\textwidth]{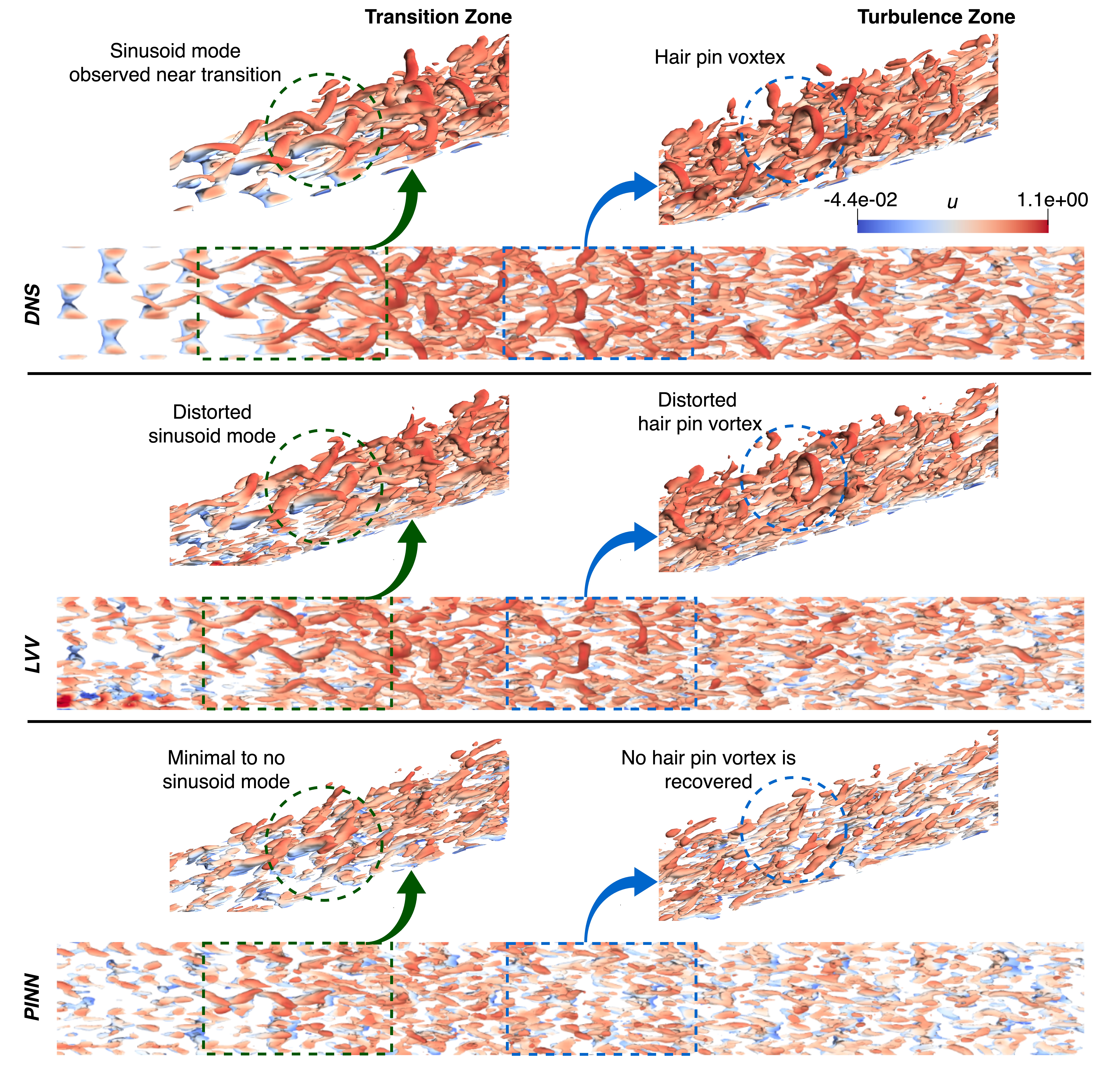}
\caption{Zoomed comparison of $Q$-criterion iso-surfaces at snapshot
$s=212$ for $P_s=8$. The same near-transition and downstream turbulent
subdomains shown in Fig.~\ref{fig:ps_4_q_zoom_s212} are extracted from DNS,
LVV, and PINN to assess the influence of increased compression on the
recovery of representative coherent structures. The near-transition box
spans $(x_{\mathrm{start}},x_{\mathrm{end}})=(1.3,1.5)$, while the
downstream turbulent box spans
$(x_{\mathrm{start}},x_{\mathrm{end}})=(1.7,1.9)$. All panels use the same
iso-surface threshold ($Q=100$), viewing direction, and streamwise-velocity
colour scale.}
\label{fig:ps_8_q_zoom_s212}
\end{figure*}

To examine local structural recovery in more detail,
Figs.~\ref{fig:ps_4_q_zoom_s212} and \ref{fig:ps_8_q_zoom_s212} present
zoomed views of two representative subdomains at snapshot $s=212$. The first
subdomain is selected from the near-transition region, where the DNS field
shows a developing sinusoid vortical structure. The second subdomain is
selected farther downstream in the turbulent region, where the DNS field
contains a denser population of interacting vortical structures and a
representative hairpin-like vortex. These two subdomains are chosen to assess
whether the reconstructed fields preserve both the instability-related
structure near transition and the more developed vortical topology in the
turbulent region.

For $P_s=4$, LVV recovers both local features with good qualitative agreement
with DNS. In the near-transition region, the sinusoid vortical structure is
retained with similar orientation and continuity. In the downstream turbulent
region, the hairpin-like vortical structure is reproduced in LVV, retaining its identifiable geometry. PINN captures the presence of rotation-dominated structures, but the sinusoid structure is more distorted and the hairpin-like vortex is less continuous than in LVV.

For $P_s=8$, the effect of increased sparsity becomes more evident. In the
near-transition region, LVV retains only a distorted representation of the
sinusoid structure, while PINN shows minimal to no recovery of this feature. In
the downstream turbulent region, LVV preserves a distorted but still
recognisable remnant of the hairpin-like vortex, whereas PINN does not recover
a clearly identifiable hairpin-like topology. These zoomed comparisons support
the full-domain observation that coherent vortical structures deteriorate with
increasing compression, with LVV retaining greater topological fidelity than
PINN.

To complement the visual assessment, a quantitative measure of coherent
vortical content is defined as
\begin{equation}
\phi_Q=\frac{V(Q>Q_0)}{V_{\mathrm{domain}}},
\end{equation}
where $Q_0=100$ is the iso-surface threshold used in
Figs.~\ref{fig:q_iso_surfaces_lvv_pinn}, \ref{fig:ps_4_q_zoom_s212}, and
\ref{fig:ps_8_q_zoom_s212}. Here, $\phi_Q$ denotes the fraction of the domain
occupied by rotation-dominated structures. To assess sensitivity to the
threshold choice, additional analyses were performed for $Q_0=50$ and
$Q_0=150$. These analyses showed qualitatively similar trends to those
obtained for $Q_0=100$; therefore, only the results for $Q_0=100$ are
reported.

\begin{table}[t]
\centering
\setlength{\tabcolsep}{10pt}
\begin{tabular}{cccccc}
\toprule
\textbf{Snapshot} & \textbf{DNS} &
\multicolumn{2}{c}{$P_s=4$} &
\multicolumn{2}{c}{$P_s=8$} \\
\cmidrule(lr){3-4} \cmidrule(lr){5-6}
& & \textbf{LVV} & \textbf{PINN} & \textbf{LVV} & \textbf{PINN} \\
\midrule
212 & 0.143 & 0.140 & 0.121 & 0.126 & 0.0860 \\
478 & 0.164 & 0.157 & 0.132 & 0.131 & 0.0889 \\
\bottomrule
\end{tabular}
\caption{Comparison of the fraction of domain points with $Q>100$ for DNS,
LVV, and PINN at snapshots 212 and 478 for $P_s=4$ and $P_s=8$.}
\label{tab:q_fraction}
\end{table}

Table~\ref{tab:q_fraction} reports the vortical-volume fraction $\phi_Q$ for
DNS, LVV, and PINN. At $P_s=4$, LVV remains close to the DNS reference; for
example, at $s=212$, $\phi_Q=0.140$ for LVV compared with $0.143$ for DNS.
PINN also retains a substantial fraction of rotation-dominated regions, but
consistently underestimates $\phi_Q$, indicating reduced vortical content.

At $P_s=8$, the reduction in $\phi_Q$ becomes more pronounced for both
methods, with a stronger decrease for PINN. For example, at $s=212$, PINN
gives $\phi_Q=0.0860$, compared with $0.143$ for DNS, whereas LVV retains
$\phi_Q=0.126$. Similar trends are obtained at $s=478$. Together with the
full-domain and zoomed visual comparisons, these values indicate that LVV
preserves a larger fraction of the rotation-dominated topology as sparsity
increases.

\subsection{Summary of Results{\label{subsec:summary_of_results}}}

The results from Reynolds stresses (Section~\ref{subsec:reynolds_stresses}), and Q-isosurfaces (Section~\ref{subsec:Qisosurface_generalization}) indicate that LVV preserves turbulent statistics better than PINN, likely due to its access to higher-resolution data. In contrast, PINN demonstrates greater stability in terms of MSE (Section~\ref{subsec:inst_vel}) and energy spectra (Section~\ref{subsec:energy_spectra}) at higher \(P_s\). Although this may seem counterintuitive, this can be explained by examining the error distribution across 77 $z$ planes. As shown in Figure~\ref{fig:mse_line_plot}, LVV exhibits increasing reconstruction errors away from the training region, contributing to higher MSE and degraded energy spectra. However, as \(z = 19\) and \(z = 38\) are outside the high-error region, LVV shows comparatively better agreement with reference turbulent statistics than PINN. PINN, in contrast, maintains a more uniform error distribution across the entire domain, exhibiting stable out-of-domain error characteristics. Thus, performance in the training domain may serve as a good measure of PINN's out-of-domain generalisation capability.

In summary, at low \(P_s\), both models recover multi-scale structures reasonably well, with LVV providing improved statistical fidelity due to its use of high-resolution labels. At high \(P_s\), performance degrades for both models: LVV deteriorates more significantly away from the training region due to overfitting, while PINN remains more stable but exhibits consistently higher errors, reflecting the difficulty of reconstructing velocity fields using physics-based constraints. Overall, the results from TKE, Reynolds stresses, and coherent-structure analysis provide a consistent picture of model performance. Large-scale flow structures and key turbulence statistics are reasonably well recovered at lower compression, whereas deviations become increasingly evident in smaller scales, turbulence statistics, and coherent vortical structures as sparsity increases. This combined assessment highlights the trade-off between compression, reconstruction accuracy, and held-out extrapolation behaviour, and provides the basis for discussing the broader significance of the present work.

\section{Implications for flow reconstruction and data compression}
The present results have important implications and broader significance for both data-driven flow reconstruction and turbulence data compression. First, the results show that supervised super-resolution can serve not only as a tool for reducing storage requirements but also as a viable approach for reconstructing turbulent flow fields while preserving important physical features, including spectral content, turbulence statistics, and coherent structures. This suggests that compression strategies based on learned reconstruction may be designed using physical fidelity, rather than point-wise error alone, as a guiding criterion.

Second, the results highlight the complementary roles of supervised and physics-informed approaches. While the vorticity-augmented supervised model (LVV) provides higher reconstruction fidelity at lower $P_s$ when high-resolution training data are available, the physics-informed model indicates the potential to reconstruct turbulent flow from sparse measurements without high-resolution labels. This distinction has implications for applications where labelled data are limited, such as experimental reconstruction, reduced sensing, and inverse flow estimation.

Third, by considering roughness-induced three-dimensional transitional turbulence and implementing a partially assisted compressible PINN formulation based on the three-dimensional unsteady compressible Navier--Stokes equations, the study extends current flow-reconstruction investigations beyond simplified canonical settings. Taken together, these results suggest the broader potential of combining data-driven and physics-informed learning for compression, sparse reconstruction, and, ultimately, data-enabled modelling of compressible turbulent flows.

\section{Conclusion and Future Work}
\label{sec:conclusion}
This work investigated the reconstruction of roughness-induced transitional turbulence from sparse data using supervised and physics-informed approaches, with an emphasis on recovering not only pointwise accuracy but also spectral content, turbulence statistics, and coherent structures. Based on the analyses performed for compression levels $P_s=4$ and $8$, the principal conclusions are as follows:

\begin{enumerate}

\item[(i)] At lower compression ($P_s=4$), both LVV and PINN recover dominant large-scale flow structures and key turbulence statistics with reasonable fidelity, although LVV generally shows closer agreement with DNS.

\item[(ii)] As compression increases to $P_s=8$, reconstruction fidelity degrades first in the smaller scales, followed by increasing deviations in turbulence statistics and coherent vortical structures, highlighting the increasing difficulty of recovering fine-scale dynamics under sparse sampling.

\item[(iii)] 
The physics-informed model (PINN) indicates the potential to reconstruct turbulent flow from sparse measurements without high-resolution labels and exhibits comparatively stable held-out extrapolation behaviour under increased sparsity.

\item[(iv)] Overall, the present study suggests that super-resolution can serve not only as a tool for data compression, but also as an approach for physically informed reconstruction of turbulent flows from sparse data.
\end{enumerate}
While the proposed frameworks yield encouraging results, several limitations in the current study remain to be addressed. First, the quantitative evaluation was constrained to only two discrete down-sampling factors ($P_s=4$ and $8$). Second, the super-resolution models were restricted to reconstructing the three velocity components ($u, v, w$), relying on reference data to supply the thermodynamic fields. 


Consequently, future research directions will focus on reconstructing all primitive variables simultaneously, improving derivative evaluation within the physics-informed formulation, exploring higher compression levels and more challenging flow regimes. Ultimately, the goal is to develop physics-informed neural network solvers for unsteady three-dimensional compressible flows using only the 3D compressible NSE, initial conditions, and boundary conditions.

\section{Appendix}
\begin{table*}[t]
\centering
\small
\begin{tabular}{@{}llSSSc@{}}
\toprule
\textbf{Model} & \textbf{Dataset} &
\textbf{nMSE($u$)} & \textbf{nMSE($v$)} & \textbf{nMSE($w$)} &
\textbf{$\boldsymbol{\mathcal{E}^*_{\text{spec}}}$} \\
\midrule

\multirow{2}{*}{S}   
& Training   & 6.06e-4 & 2.77e-2 & 2.79e-2 & \multirow{2}{*}{$2.28\times10^{-2}$} \\
& Validation & 6.22e-4 & 3.05e-2 & 3.01e-2 & \\
\midrule

\multirow{2}{*}{VS}  
& Training   & 6.07e-4 & 2.55e-2 & 3.07e-2 & \multirow{2}{*}{$2.19\times10^{-2}$} \\
& Validation & 6.24e-4 & 2.81e-2 & 3.27e-2 & \\
\midrule

\multirow{2}{*}{LVV} 
& Training   & 4.35e-4 & 2.99e-2 & 3.28e-2 & \multirow{2}{*}{$2.41\times10^{-2}$} \\
& Validation & 4.47e-4 & 3.28e-2 & 3.53e-2 & \\
\bottomrule
\end{tabular}

\caption{Ablation study for supervised models at $P_s=4$ and across  77 $z$ planes, comparing velocity-only (S), velocity-gradient augmented (VS), and LVV training across training and validation snapshots}
\label{tab:nmse_ps4_stacked}
\end{table*}
\subsection{Implementation of Navier-Stokes loss {\label{subsec:appendix_nvlos}}}
The purpose of this section is to discuss the implementation details of the Navier-Stokes loss used in training in Section \ref{subsec:NVloss}. The governing three-dimensional compressible and unsteady Navier--Stokes equations in Cartesian coordinates are written as
\begin{align}
\frac{\partial Q}{\partial t} + 
\frac{\partial E}{\partial x} + 
\frac{\partial F}{\partial y} + 
\frac{\partial G}{\partial z} =
\frac{\partial E_{v}}{\partial x} + 
\frac{\partial F_{v}}{\partial y} + 
\frac{\partial G_{v}}{\partial z},
\label{eq:compressible_NS}
\end{align}
where the conserved variable vector $Q$ and flux vectors $(E,F,G,E_v,F_v,G_v)$ are defined as
\begin{align*}
&Q = [\rho,\rho u,\rho v,\rho w,\rho e_t]^T, \\
&E = [\rho u, \rho u^2 + p, \rho uv, \rho wu, (\rho e_t + p)u]^T, \\
&F = [\rho v, \rho uv, \rho v^2 + p, \rho wv, (\rho e_t + p)v]^T, \\
&G = [\rho w, \rho uw, \rho vw, \rho w^2 + p, (\rho e_t + p)w]^T, \\
&E_v = [0, \tau_{xx}, \tau_{xy}, \tau_{xz}, u\tau_{xx} + v\tau_{xy} + w\tau_{xz} + q_x]^T, \\
&F_v = [0, \tau_{yx}, \tau_{yy}, \tau_{yz}, u\tau_{yx} + v\tau_{yy} + w\tau_{yz} + q_y]^T, \\
&G_v = [0, \tau_{zx}, \tau_{zy}, \tau_{zz}, u\tau_{zx} + v\tau_{zy} + w\tau_{zz} + q_z]^T.
\end{align*}

The viscous stress components and heat flux are defined as
\begin{align*}
&\tau_{xx} = \frac{1}{Re_\infty}\Big(2\mu \frac{\partial u}{\partial x} + \lambda \nabla \cdot \vec{V}\Big), \quad
\tau_{yy} = \frac{1}{Re_\infty}\Big(2\mu \frac{\partial v}{\partial y} + \lambda \nabla \cdot \vec{V}\Big), \\
&\tau_{zz} = \frac{1}{Re_\infty}\Big(2\mu \frac{\partial w}{\partial z} + \lambda \nabla \cdot \vec{V}\Big), \quad
\tau_{xy} = \tau_{yx} = \frac{\mu}{Re_\infty}\Big(\frac{\partial u}{\partial y} + \frac{\partial v}{\partial x}\Big), \\
&\tau_{yz} = \tau_{zy} = \frac{\mu}{Re_\infty}\Big(\frac{\partial v}{\partial z} + \frac{\partial w}{\partial y}\Big), \quad
\tau_{xz} = \tau_{zx} = \frac{\mu}{Re_\infty}\Big(\frac{\partial w}{\partial x} + \frac{\partial u}{\partial z}\Big), \\
&q_x = \frac{\mu}{(\gamma-1) Pr Re_\infty M_\infty^2} \frac{\partial T}{\partial x}, \quad
q_y = \frac{\mu}{(\gamma-1) Pr Re_\infty M_\infty^2} \frac{\partial T}{\partial y}, \quad\\
&q_z = \frac{\mu}{(\gamma-1) Pr Re_\infty M_\infty^2} \frac{\partial T}{\partial z}.
\end{align*}
where $\vec{V}=(u,v,w)$ denotes the velocity vector.
Since the dataset is defined on a non-uniform grid, spatial derivatives are evaluated using the grid metrics. The derivative of a generic scalar field $f$ is computed as
\begin{equation}
\begin{bmatrix}
\displaystyle \frac{\partial f}{\partial x} \\[0.5em]
\displaystyle \frac{\partial f}{\partial y} \\[0.5em]
\displaystyle \frac{\partial f}{\partial z}
\end{bmatrix}
=
\begin{bmatrix}
\displaystyle \frac{\partial f}{\partial i}\frac{\partial i}{\partial x} + \frac{\partial f}{\partial j}\frac{\partial j}{\partial x} + \frac{\partial f}{\partial k}\frac{\partial k}{\partial x} \\[0.5em]
\displaystyle \frac{\partial f}{\partial i}\frac{\partial i}{\partial y} + \frac{\partial f}{\partial j}\frac{\partial j}{\partial y} + \frac{\partial f}{\partial k}\frac{\partial k}{\partial y} \\[0.5em]
\displaystyle \frac{\partial f}{\partial i}\frac{\partial i}{\partial z} + \frac{\partial f}{\partial j}\frac{\partial j}{\partial z} + \frac{\partial f}{\partial k}\frac{\partial k}{\partial z} 
\end{bmatrix}.
\label{eq:derivative_calc}
\end{equation}

In the present implementation, the spatial derivatives required for $T$ are computed using a fourth-order explicit finite-difference scheme~\cite{gaitonde1998high}. The resulting Navier--Stokes residuals are assembled into the physics-based loss during training. The current implementation of PINN approximates the exact pointwise gradient ($\frac{\partial f_i}{\partial x_i}$) by the total spatial sensitivity of the grid ($\sum_{j \in \mathcal{R}(i)} \frac{\partial f_j}{\partial x_i}$), where $\mathcal{R}(i)$ is the receptive field of the network at point $i$, induced by its convolutional operations. This is introduced to avoid the prohibitive cost of computing a full Jacobian over a large 3D domain, trading strictly localised differentiation for an efficient, volume-wide aggregated gradient during the backward pass.

\subsection{Contribution of Vorticity-based penalty}

To assess the contribution of the vorticity-based penalty, an ablation study is performed comparing velocity-only (S)(refer to equation ~\ref{eq:loss_super}), Vorticity scaling (VS) (Equation ~\ref{eq:loss_VS}), and LVV training.
\begin{equation}
J_{S} =
\frac{1}{b_sN_sc}\sum_{b=1}^{b_s}\sum_{i=1}^{N_s}
\left\|\hat{\mathbf{f}}_{b,i} - \mathbf{f}_{b,i}\right\|_2^2 ,
\label{eq:loss_super}
\end{equation}
\begin{equation}
\begin{split}
J_{\mathrm{VS}} &=
\frac{1}{b_s N_s c}
\sum_{b=1}^{b_s}\sum_{i=1}^{N_s}
\Biggl[
1.0 + \max\Biggl(
10^{-2}, \\
&\qquad\min\Biggl(
1.0,
\frac{\|\boldsymbol{\omega}_{b,i}\|_2}
{\max\limits_{j\in\{1,\dots,N_s\}}
\|\boldsymbol{\omega}_{b,j}\|_2}
\Biggr)
\Biggr)
\Biggr]
\|\hat{\mathbf{f}}_{b,i}-\mathbf{f}_{b,i}\|_2^2 .
\end{split}
\label{eq:loss_VS}
\end{equation}
Here, $\mathbf{f}_{b,i}$ and $\hat{\mathbf{f}}_{b,i}$ represent the true and predicted velocity vectors and $\omega$ denotes true vorticity.  VS is similar to the weighted mean-squared approach, where we scale the error at each point by the magnitude of the true vorticity. In VS, the true vorticity values are rescaled to the range ([1,2]) to improve numerical stability during training.

 We compare the performance of three models using nMSE (defined in equation \ref{eq:nmse}) and spectral reconstruction error. The scalar measure of spectral reconstruction error is defined as
\begin{equation}
\mathcal{E}^{*}_{\mathrm{spec}} =
\frac{\int \left| E_{\mathrm{pred}}(k) - E_{\mathrm{DNS}}(k) \right| \, dk}
{\int E_{\mathrm{DNS}}(k) \, dk},
\label{eq:spec_error}
\end{equation}
where $E_{\mathrm{pred}}(k)$ and $E_{\mathrm{DNS}}(k)$ denote the reconstructed and reference energy spectra, respectively. This normalized quantity provides a compact assessment of spectral fidelity across all resolved scales. Table~\ref{tab:nmse_ps4_stacked} summarises the corresponding reconstruction errors and spectral metrics at $ P_s = 4$. The results indicate that differences among the three supervised formulations are relatively small for the present dataset, with all models exhibiting comparable reconstruction accuracy. LVV shows slightly improved performance in the primary velocity component, while VS yields marginally lower spectral error. Based on this overall assessment, LVV is adopted for all subsequent supervised reconstruction results for brevity.

\section{Acknowledgements}
The authors acknowledge the National Supercomputing Mission (NSM) for providing computing resources of ‘PARAM SHAKTHI’ at the P. G. Senapathy Centre, Play Field Avenue, Indian Institute of Technology Madras, Chennai, Tamil Nadu 600036, which is implemented by the Centre for Development of Advanced Computing (C-DAC) and supported by the Ministry of Electronics and Information Technology (MeitY) and the Department of Science and Technology (DST), Government of India. Additionally, we acknowledge the National PARAM Supercomputing Facility (NPSF) for providing computing resources on the PARAM Siddhi cluster under NSM. We also thank C-DAC for their support through hackathons.

\section*{Data Availability Statement}
The data that supports the findings of this study are available from the corresponding author upon reasonable request.

\nocite{*}
\bibliographystyle{plain}  
\bibliography{citation}
\end{document}